\documentclass[aps,prl,showpacs,onecolumn,11pt]{revtex4-1}
\usepackage{graphicx}
\usepackage{graphics}
\usepackage{subfigure}
\usepackage{epstopdf}
\usepackage[T1]{fontenc}
\usepackage[latin1]{inputenc}
\usepackage{graphics}
\usepackage{amsthm}
\usepackage{amsfonts}
\usepackage{amsmath}
\usepackage{amssymb}
\usepackage{amsthm}
\usepackage{color}
\providecommand{\LyX}{L\kern-.1667em\lower.25em\hbox{Y}\kern-.125emX\@}
\makeatother

\begin{document}
\title{Toward robust phase-locking in {\em Melibe} swim central pattern generator models}

\author{Sajiya Jalil, Dane Allen, Joe Youker and Andrey Shilnikov}
\address{Neuroscience Institute and Department of Mathematics and Statistics, \\
Georgia State University, GA 30303, USA
 }

\begin{abstract}
Small groups of interneurons, abbreviated by CPG for central pattern generators, are arranged into neural networks to generate a variety of core bursting rhythms with specific phase-locked states, on distinct time scales, that govern vital motor behaviors in invertebrates such as chewing, swimming, etc. These movements in lower level animals mimic motions of organs in higher animals due to evolutionarily conserved mechanisms. Hence, various neurological diseases can be linked to abnormal movement of body parts that are regulated by a malfunctioning CPG. In this paper, we, being inspired by recent experimental studies of neuronal activity patterns recorded from a swimming motion CPG of the sea slug  {\it Melibe leonina}, examine a mathematical model of a 4-cell network that can plausibly and stably underlie the observed bursting rhythm.  We develop a dynamical systems framework for explaining the existence and robustness of phase-locked states in activity patterns produced by the modeled CPGs. The proposed tools can be used for identifying core components for other CPG networks with reliable bursting outcomes and specific phase relationships between the interneurons. Our findings can be employed for identifying or implementing the conditions for normal and pathological functioning of basic CPGs of animals and artificially intelligent prosthetics that can regulate various movements.  
\end{abstract} 
%\pacs {05.45.Xt, 87.19.La}

%\date{\today}

\draft \maketitle

\newpage
{\bf Many abnormal neurological phenomena are perturbations of normal functions of the underlying mechanisms governing the animal behaviors, specifically movements. Repetitive behaviors are often associated hypothetically with the phenomenon of rhythmogenesis in small networks that are able autonomously to generate or continue, after induction a variety of activity patterns without further external input, abrupt or not. The goal of this modeling study is to identify decisive components of a biologically based CPG that has been linked to a specific motion in a lower order animal, sea slug Melibe leonina, which produces a specific bursting pattern, as well as to identify components that ensure the pattern's robustness. 
Due to the recurrent nature of such bursting patterns of self-sustained activity we employ Poincare return maps defined on phases and phase-lags between burst initiations in the interneurons to study quantitative and qualitative properties of CPG rhythms and corresponding attractors. \\
The proposed approach is specifically tailored for various studies of neural networks in neuroscience, computational and experimental. Development of such tools and our understanding of such CPGs can be applied to gain insight into governing principles of neurological phenomena in higher order animals and can aid in treating anomalies associated with neurological disorders.}

\newpage

\section{Introduction} \label{intro}

A central pattern generator (CPG) is a neural network of a small group of synaptically coupled interneurons that is able to generate single or multiple rhythmic  outcomes without external [sensory] feedback \cite{MC96,Myoutube1,Myoutube2}. CPGs establish and govern various motor behaviors of animals such as swimming, crawling, walking etc \cite{CPG,KCF05}.  In addition, the mechanisms of such motions are evolutionarily conserved and can be related to rhythmic motions of various body parts, such as heart, lungs, legs etc, of higher order animals. Such rhythmic outcomes, often viewed as bursting patterns, can be differentiated by several timing properties, including  specific and robust  phase-locked states between   well orchestrated interneurons within the specific CPG of a particular animal.  As such, the behavior controlled by the CPG can be disrupted, or halted, after a component neuron of the network is blocked or temporarily inhibited. This would indicate that the rhythmic outcome results from synergetic interactions of all contributing interneurons, which may not be necessarily endogenous bursters, when isolated from the network. The robustness of a rhythmic outcome is an essential property allowing the CPG to  withstand or recover from perturbations, a lack of which could be the expression of various neurological diseases and disorders. 
 
Identification and modeling of a CPG underlying an animal behavior is a real challenge due to a number of factors. The realization of a behavior may require components, other than interneurons, such as synapses, which can be fast and slow,  inhibitory, excitatory, or electrical, etc. Furthermore, interneurons, which are networked within the CPG for one behavior, may contribute to another behavior as well, i.e. be multifunctional \cite{BK08,SGB08}. A whole CPG can be also multifunctional if it governs more than one behavior, in contrast to a dedicated CPG which is arranged for the purpose of a  single locomotion. Modeling studies, mathematical and computational, have proven to be useful for gaining insights into operational principles of  CPGs \cite{M87,Kopel88,SKM94,GSBC99,BS08,SHWG10}, in particular, multifunctional ones \cite{CBCB99,WCS11,Jyoutube}. This study has been inspired by recent experimental studies of neuronal activity patterns recorded from the identified interneurons 
of a [dedicated] CPG  governing the swim pattern of a sea slug  {\it Melibe leonina} \cite{SNLK11,SK11,NSLGK012} and possible consequences of understanding neurological phenomena in general.  

The occurrence of phase-locked bursting patterns is naturally observed in the voltage traces  simultaneously recorded from a few interneurons, which are argued to belong to the same CPG. In such recordings, specific delays between initiations of the active (spiking) phases of the bursting interneurons are well maintained. It is unclear so far whether the delays or phase-lags between bursting interneurons are essential for 
the CPG. An argument supporting the latter hypothesis is that the same  phase-lags 
between burst initiations have been recorded in both juvenile and adult animals \cite{SK11},  while the frequency and hence delays between burst initiations in voltage traces can vary substantially during animals' life spans.   
Bursting neurons display multiple action potentials during their active phases of bursting and remain hyper-polarized during their inactive phases. The bursts of action potentials or spikes correlate with neurotransmitters' release that allows the neurons to interact. Hence, the specific delays between the bursting patterns can be meaningful to and explanatory for CPG formation mechanisms \cite{ADGOPPS,ADOPS98,JR99,WNT02,NW02,TW05,M12}.

The question of whether a neuron belongs to a CPG is not easy. Potentially it can be resolved by linking the sequence of bursting delays with some timescale of movements during the behavior. An alternative is a perturbation approach when a targeted neuron is temporarily injected with a polarized current to trace down a metamorphic effect or a qualitative change on the network. A modeling study of the CPG, however, remains a practical approach for singling out particular features among networks that are vital for its own proper functioning \cite{CBCB97,DRR09}. In this paper, we will start with the consideration of rather idealized, symmetric configurations of CPG networks whose repetitive bursting patterns or rhythmic outcomes are known a priori. By doing so, we will probe tools developed for uncharacteristic dynamics, including a possible co-existence of several patterns in other CPG configurations. Next, we introduce, gradually, variations in the parameters to match our findings with the recorded electrophysiology of the animal. The strategy should allow us to find a minimal wiring for synaptic connections that gives rise to (robust) neuronal dynamics observed in experimental studies.

We begin by introducing the dynamical toolkit in the Methods section that is used throughout this study. It is followed by a section on a CPG example made of uncoupled half-center oscillators (HCOs). A HCO is composed of two bilaterally symmetric cells reciprocally inhibiting each other to produce  alternating (anti-phase) bursting patterns.  Such a pair can burst in-phase too \cite{pre2010,pre2012}, when, for example, it is exogenously driven by a pre-synaptic interneuron of the network \cite{BS08}.  We will consider two cases of uncoupled cs, homogeneous and heterogeneous. By comparing their dynamics, we identify the conditions giving rise to robust and unique phase-locked pattern formations, such as ones  recorded in the experiments. 
As a next step, we introduce additional synaptic connections, which are arguably known to exist in the circuitry model of the swim CPG and study their roles in regulating the mathematical models of CPG. Then, we can evaluate the parameter range for the network heterogeneity, 
which sustains the plausible phase-locked patterns, and show how the latter depend on the network configurations. Finally, we construct return maps for the phase-lags between both HCOs, not interneurons. This will further reduce the original problem (coupled 12 ordinary differential equations) to low-order maps to study synchrony, stability, coexistence and bifurcations of the bursting patterns.  In the Discussion section we will also address the challenges and future directions and application for the proposed dynamical framework.

%%%%%%%%%%%%%%%%%%%%%%%%%%%%%%
\section{Methods} \label{Methods}

In this study of 4-cell CPG networks, we employ a generic Hodgkin-Huxley-like  model of an endogenously bursting interneuron as an elementary block; the building blocks will be the HCO that can intrinsically burst either in-phase, in general, or anti-phase, in particular.   This reduced model, describing the dynamics of a leech heart interneuron, has been extensively studied and biophysically calibrated to  demonstrate a variety of activity patterns typical for various invertebrates \cite{ALS05,CCS07}. Depending on external drive, the interneuron model can produce tonic spiking activity, be an endogenous burster, or settle down to hyper- and depolarized quiescence states. 
The model has turned out to be very reach dynamically  as can demonstrate a number of global bifurcations at the activity transitions, like the blue sky catastrophe, bi-stability and chaos due to homoclinic saddle orbits \cite{ALS12}. In this study, individually each post-synaptic interneuron is an endogenous burster that can become temporarily shut at the hyper-polarized state by an inhibitory current originating from pre-synaptic interneuron(s) \cite{BS08,WCS11}.   

This level of accuracy is important for understanding CPG mechanisms as models must be compared with real animal behaviors for testing our hypotheses, however, eventually model parameters would be modified to fit neurological phenomena in mammals, in particular humans, for investigating disorders with neurological origins. 

Below, we will consider several types of CPGs made of the interneuron models weakly coupled 
by synapses, chemical: inhibitory and excitatory, and   electrical, referred to as a gap junction.
The equations of the coupled model are given in Appendix. Chemical synapses, inhibitory and excitatory, are described within the framework of the fast threshold modulation (FTM) paradigm \cite{FTM}, which has been proven to meet some basic conditions for coupled bursters \cite{pre2010,pre2012}. The strength of coupling is controlled 
by the maximal conductance, $g_{\rm syn}$, for the synaptic current. Besides $g_{\rm syn} \ll 1$, as its magnitude should be sufficient to guarantee  a slow rate of progression of bursting patterns, transitioning toward a phase-locked state, if any. We ensure that the convergence is not due to a symmetry of network interactions; some deviations, $\delta_{ij}$, from the nominal values are introduced in the inhibitory synapses:   $g^{\rm inh}_{ij}=g_{\rm syn}(1+\delta_{ij})$. Unless otherwise mentioned, $\delta_{12}=8 \times 10^{-3}$, $\delta_{12}=-6 \times 10^{-3}$, $\delta_{34}=9 \times 10^{-3}$, $\delta_{43}=-1 \times 10^{-2}$, $\delta_{32}=2 \times 10^{-3}$ and $\delta_{41}=-2 \times 10^{-3}$. 
 
We must point out that a transient trajectory can converge to a network attractor rather quickly even in a weakly coupled  case ($g_{\rm syn} \ll 1$). Such a quick convergence can occur when the endogenously bursting interneuron is initially close to a transition to a quiescent steady state through a slow-time scale
 bifurcation like saddle-node or homoclinic.  
 In this case, the trajectory can come close by or cross back and forth the corresponding (bifurcation) boundary when an interneuron receives (or is released from) a flux of inhibition from another pre-synaptic interneuron on the network. 
Physiologically speaking, such neuromodulation can be viewed as an analogues to the mathematical phenomena of bifurcations through perturbations.  Natural substances such as serotonin released by the animal can alter intrinsic properties of the individual neurons affecting the efficiency of a CPG and vary its temporal characteristics without breaking the bursting pattern {\em per se.} \cite{CTMKF07}. 
    
\begin{figure*}[!htbp]
\centering
  \subfigure [][] {\resizebox*{.50\columnwidth}{!}{\includegraphics{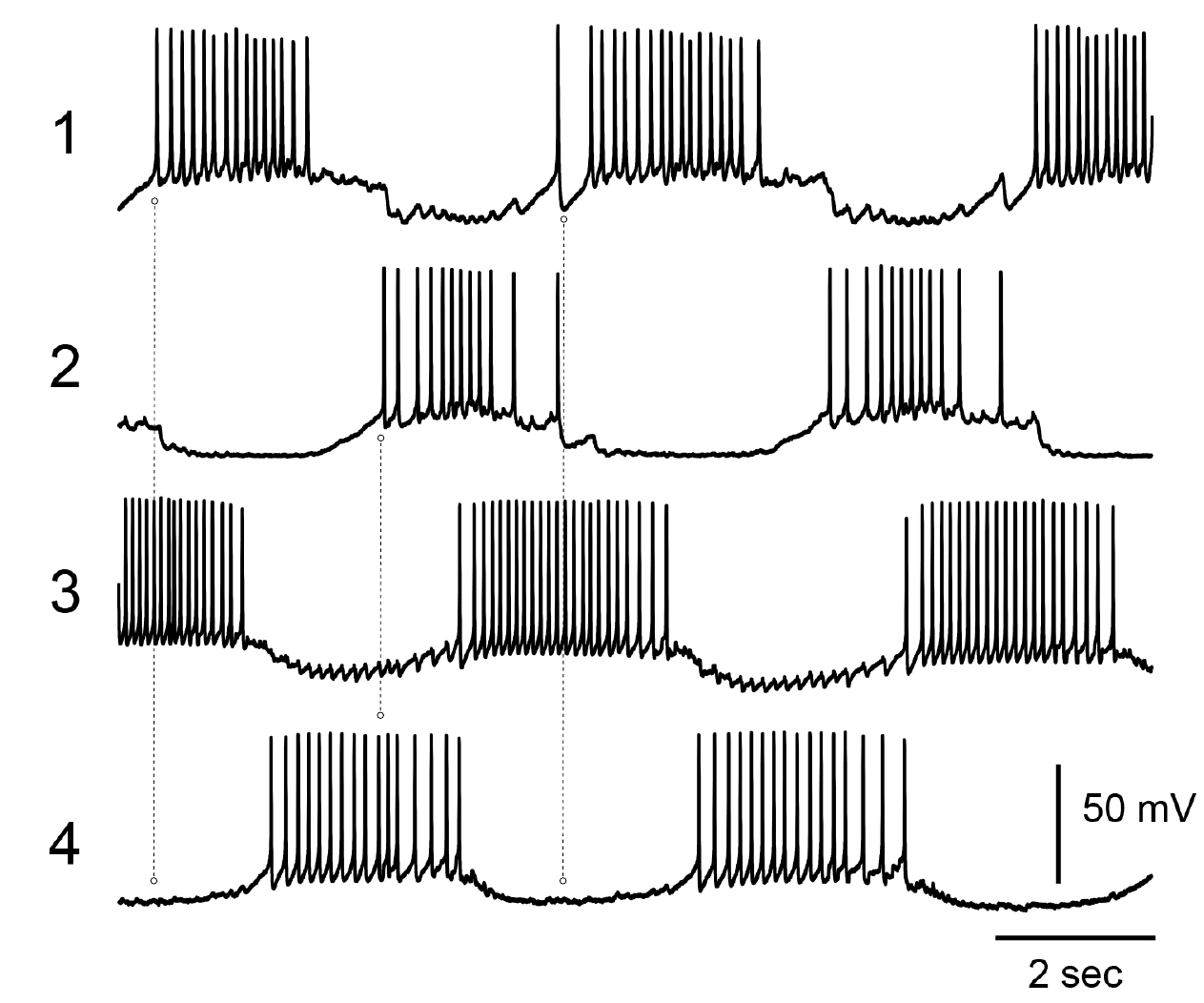}}} 
  \subfigure [][] {\resizebox*{.42\columnwidth}{!}{\includegraphics{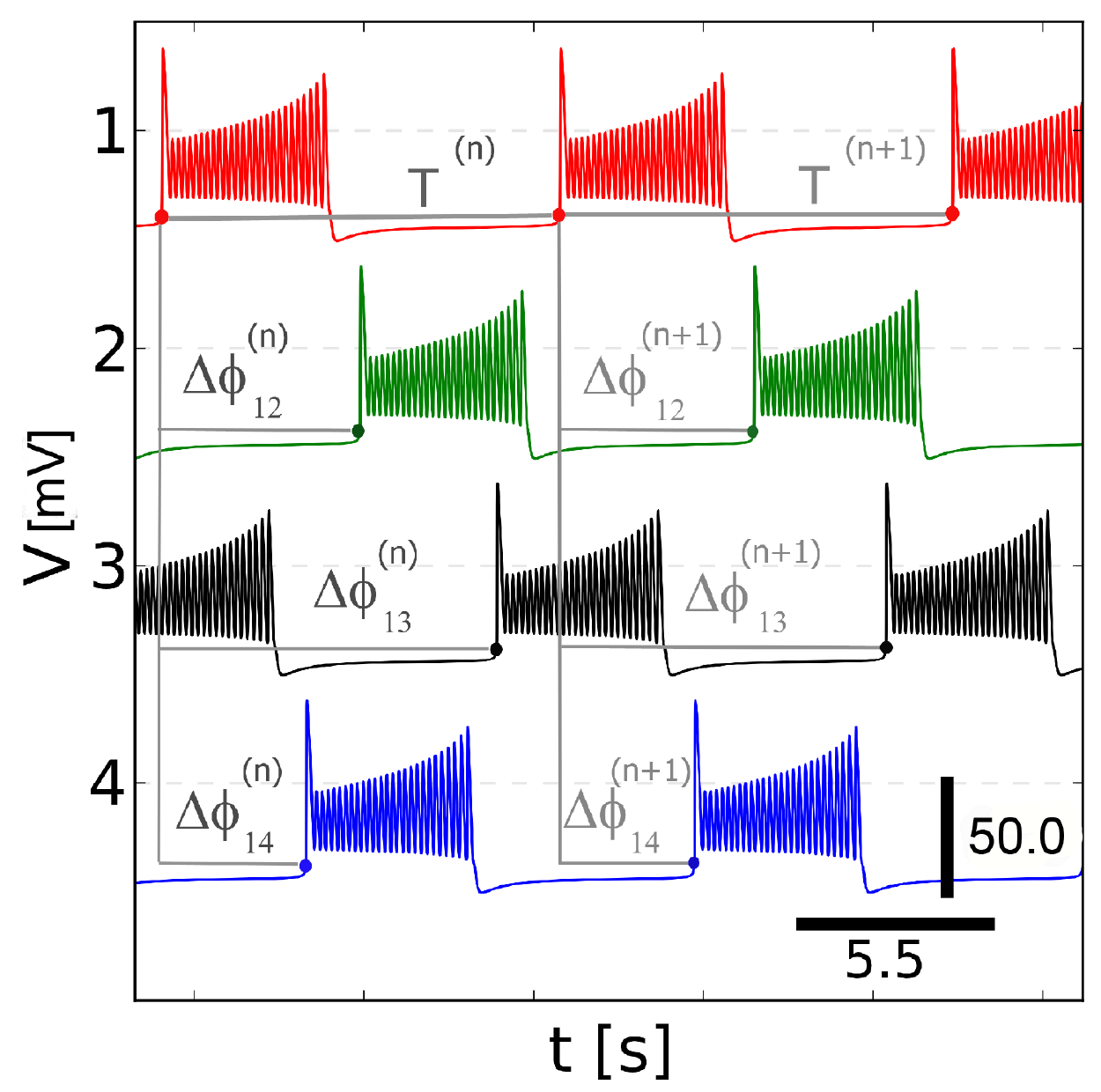}}}
     \caption{(a) Typical bursting pattern intracellularly recorded from identified interneurons of the Melibe swim CPG with 
characteristic $\frac{3}{4}$-phase shift. Recording provided courtesy of A.~Sakurai. (b)  Shifted phase-lag voltage traces generated by the CPG model. Three phase-lags, $\Delta \phi_{1j}^{(n)}$, are defined by the time delays between burst initiations (indicated by dots) of the reference interneuron~1  and the following three interneurons, 2,3 and 4, scaled over the period $T^{(n)}_1$ or recurrence time of the network.  Black bars indicate the voltage zero level and the time scale in traces.} 
\label{fig1}
\end{figure*}

In the CPG mathematical model we set all the parameters so that the individual and networked interneurons remain endogenous bursters.  The duty cycle of bursting of the interneurons, which is a fraction of the period during which the interneuron is active, persistently stays around $50\%$, i.e. the burst (active) durations and hyper-polarized (inactive) periods are almost equal.  Figure \ref{fig1} shows a typical bursting pattern that resembles the experimental recordings from four interneurons of the {\em Melibe} swim CPG.

The specific delays between the burst initiations between the interneurons are the key characteristics of each given CPG.  The core idea underlying our computational tools is inspired by  ``wet lab'' experimental observations and therefore tailored for neuroscience \cite{WCS11}. In essence, it requires only the voltage recordings from the mathematical interneurons, and therefore does not explicitly rely on the gating variables from a Hodgkin-Huxley type model. We intentionally choose the phases based on the membrane voltages, as these are  basically the only  variables that can be experimentally assessed and measured. Moreover, as in wet experiments, we have control over, and hence can maintain the initial delays, or phase distribution by releasing the interneurons from inhibition at various times after or prior of the release of the reference neuron.

The phase relationships between the coupled interneurons are defined through specific events,
 $\left \{ t^{(n)}_1,\, t^{(n)}_2,\, t^{(n)}_3, t^{(n)}_4 \right \}$, that occur when their voltages reach an auxiliary threshold, $\Theta_{th}=-0.045$V, set above the hyperpolarized voltage and below the spike oscillations.  Such events indicate the initiation of the $n^{\mathrm{th}}$ sequential  bursts in the interneurons, see Fig.~\ref{fig1}.   

\begin{figure*}[!htbp]
  \centering
  \subfigure [][] {\resizebox*{.22\columnwidth}{!}{\includegraphics{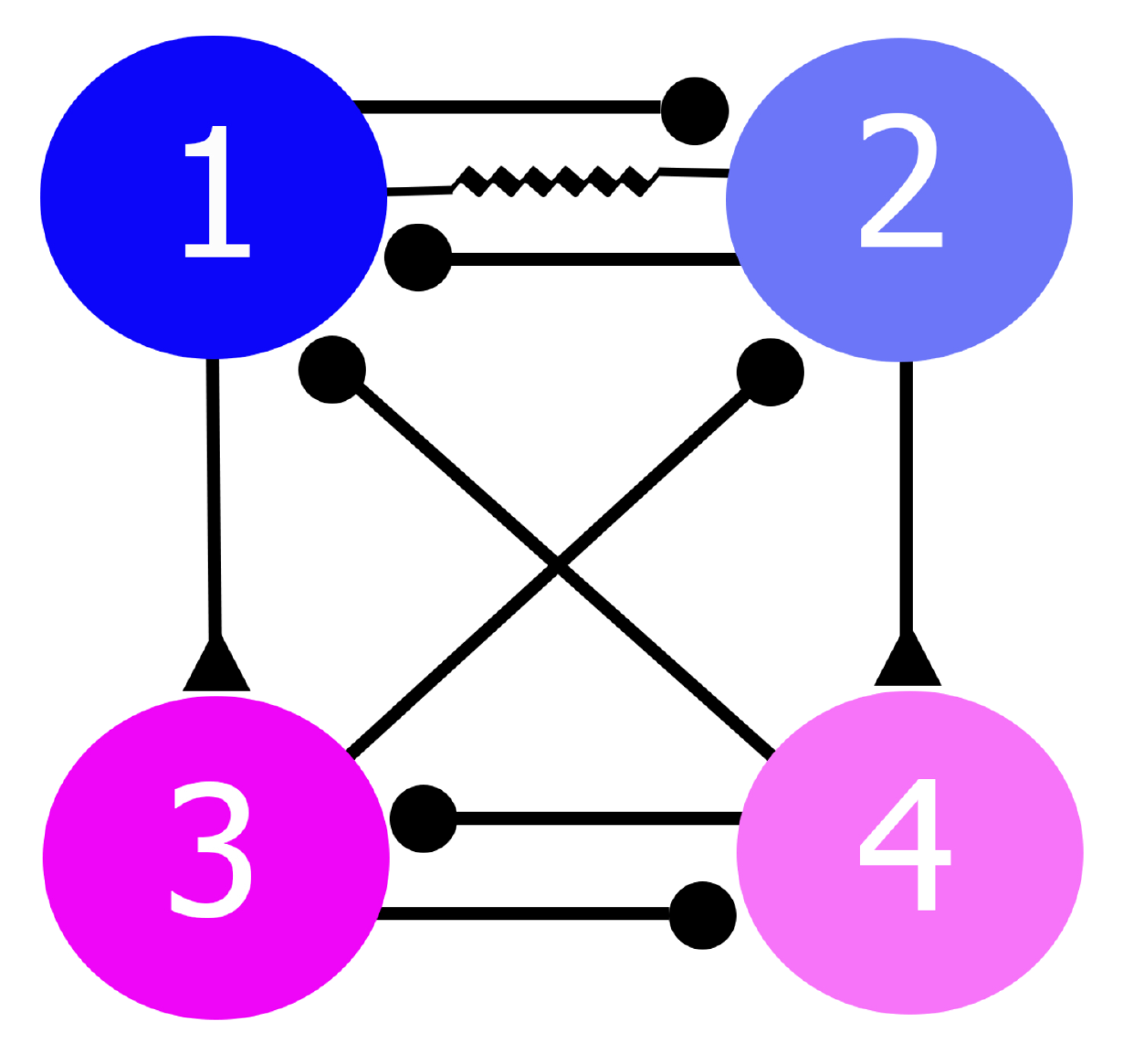}} \label{sfig3a}}
  \subfigure [][] {\resizebox*{.45\columnwidth}{!}{\includegraphics{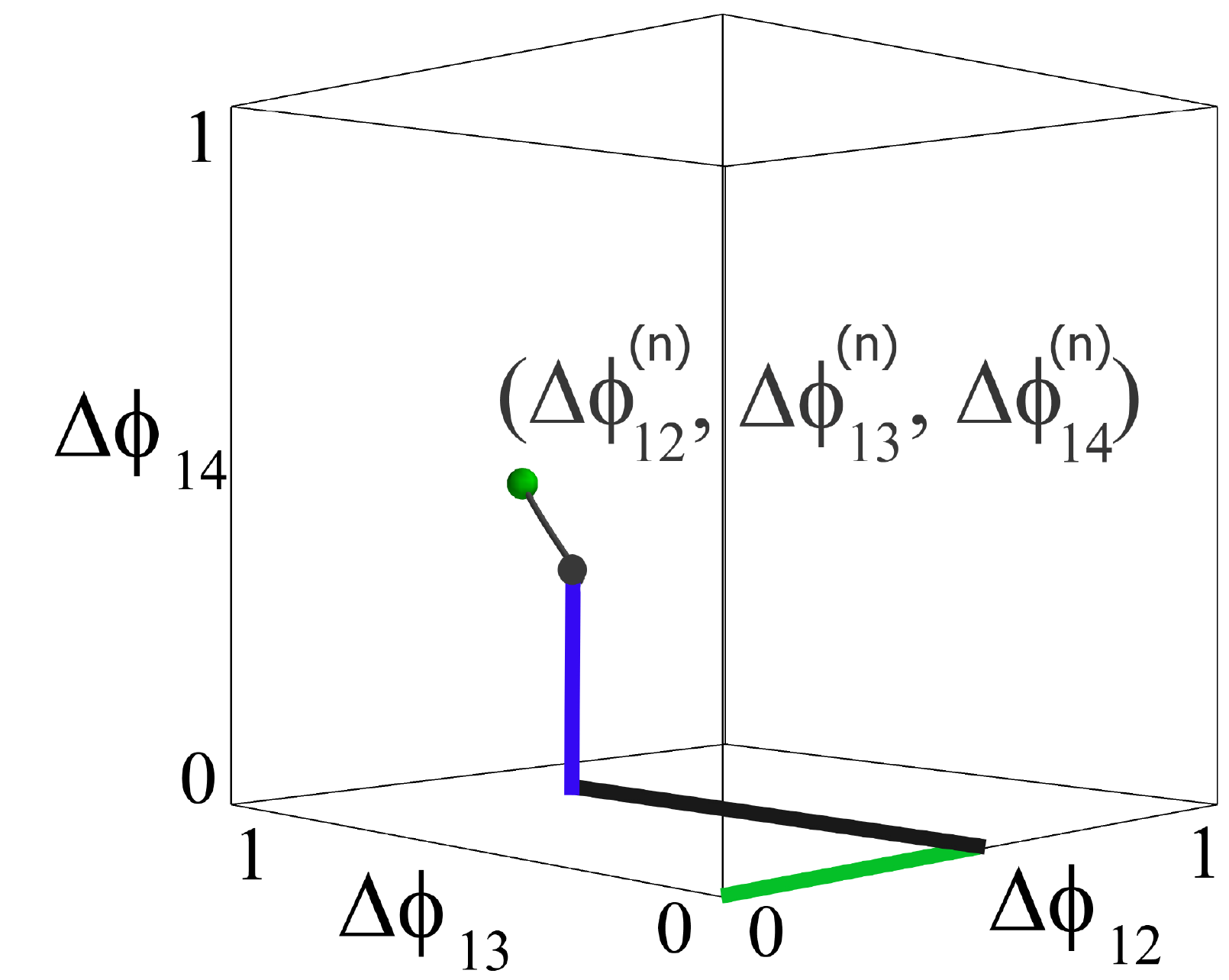}}\label{sfig3b}}\\
  \subfigure [][] {\resizebox*{.5\columnwidth}{!}{\includegraphics{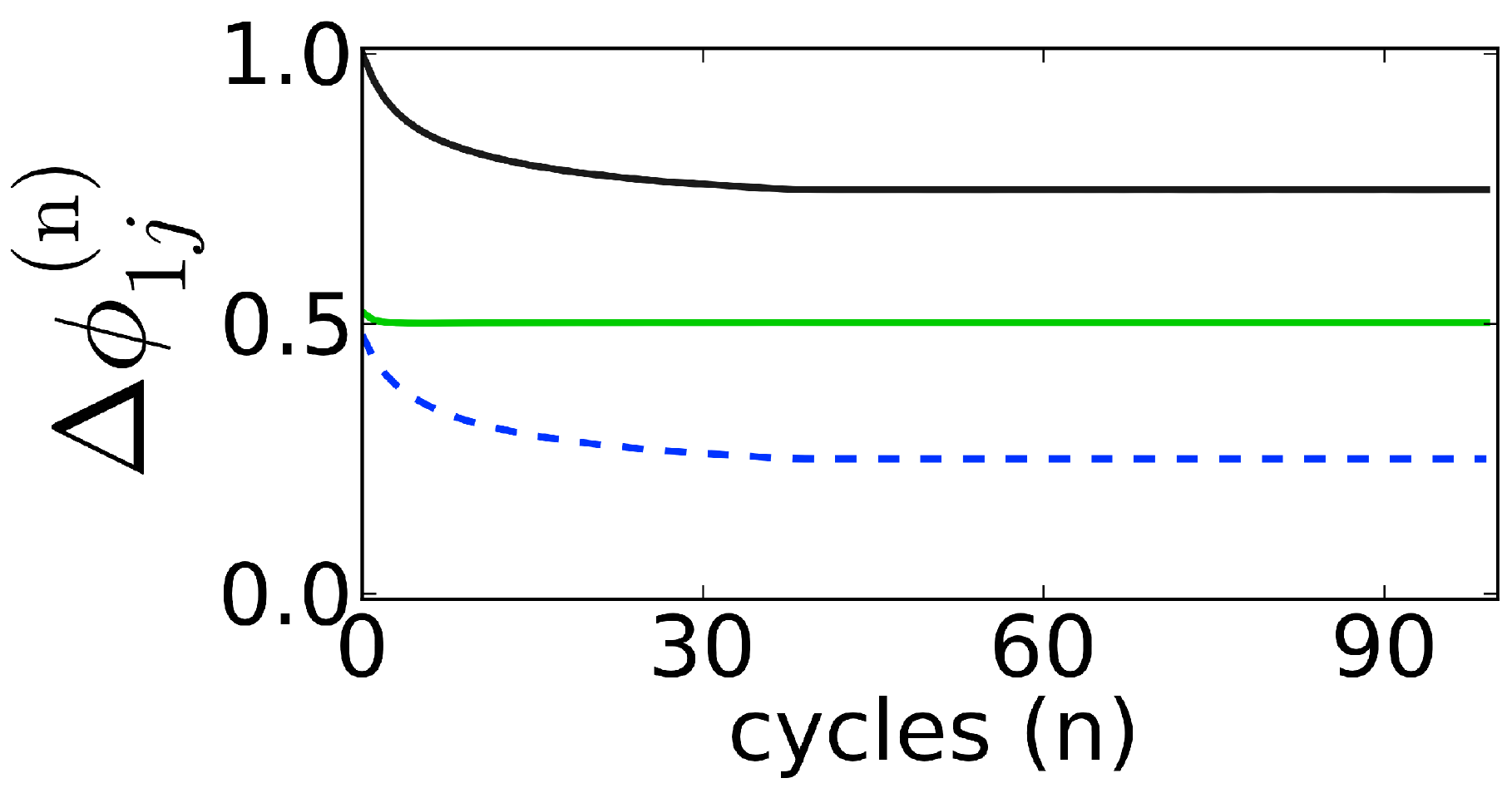}} \label{sfig3c}}
  \subfigure [][] {\resizebox*{.45\columnwidth}{!}{\includegraphics{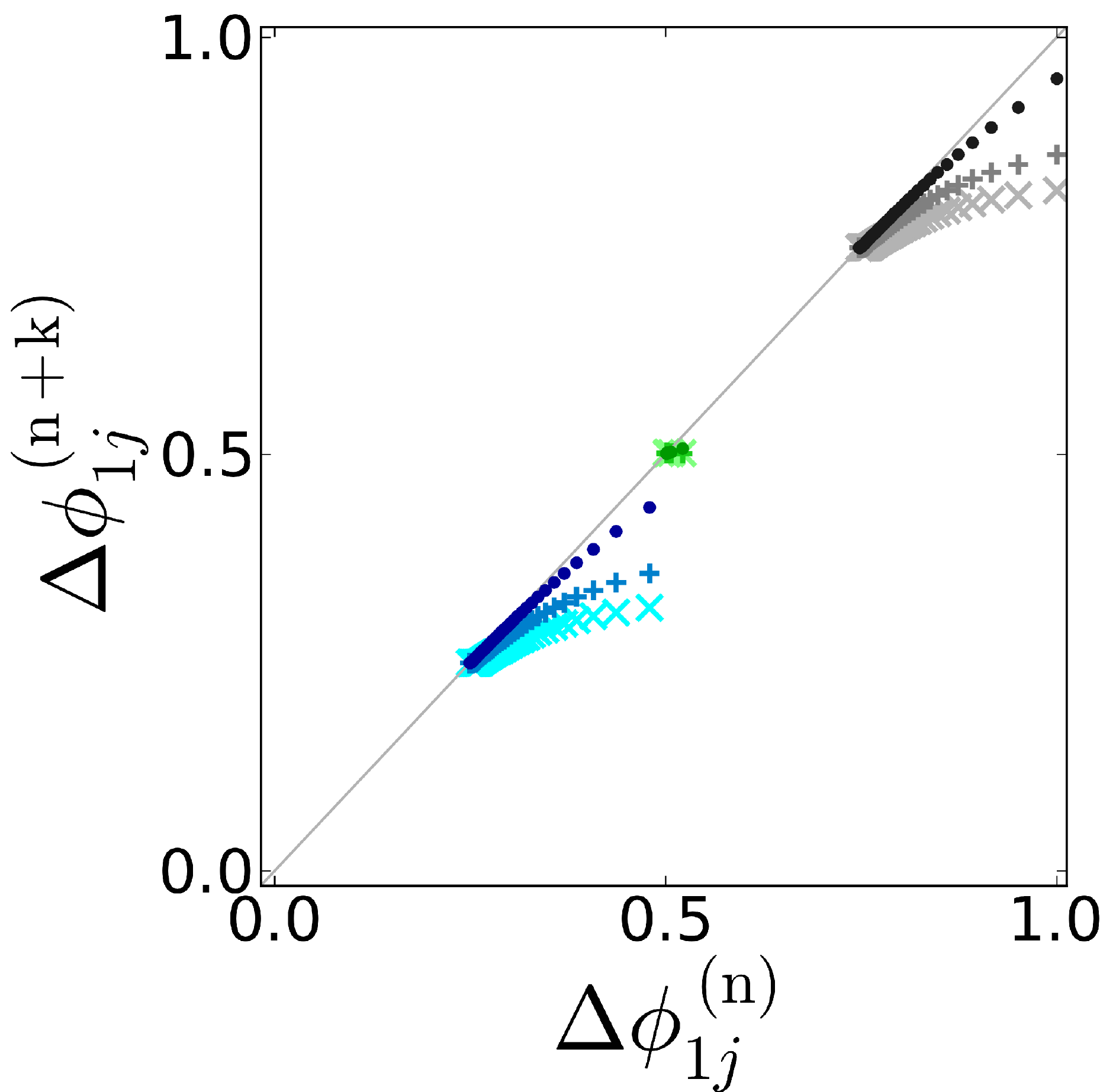}} \label{sfig3d}}
  \caption{ (a) Schematic diagram of the swim CPG with two dedicated HCOs made of couples:  Si1/2-L and Si1/2-R, and  Si3-L and Si3-R. 
Dots, $\bullet$, solid triangles, $\blacktriangle$, and the resistor represent inhibitory, excitatory and electrical synapses of conductance strengths $g_{ij}^{\rm inh}$, $g_{ij}^{\rm exc}$ and $g_{ij}^{\rm elec}$, respectively. 
 (b) Phase-lag tuple, ($\Delta \phi_{12}^{(n)}, \Delta \phi_{13}^{(n)}, \Delta \phi_{14}^{(n)}$), is a phase point on the trajectory 
in the unit cube for 3D torus. Green sphere indicates an initial phase-lags tuple; green, black and blue lines indicate $n$-th $\Delta \phi_{12}$, $\Delta \phi_{13}$, and $\Delta \phi_{14}$ coordinates of the trajectory, where $n$ varies from $0$ to $100$ along the depicted trajectory. 
 (c)  Evolutions of the phase-lags, $\Delta \phi_{12}^{(n)}$ (green), $\Delta \phi_{13}^{(n)}$ (black) and $\Delta \phi_{14}^{(n)}$ (blue) 
plotted against the burst cycle, $n$.   (d) 1D Poincar\'e maps of degrees: $k=1$ (dots), $k=5$ (small dark pluses), $k=10$ (large light crosses))  for the (color-coded) phase-lags showing the convergence to fixed  points on the $45^\circ$-line: $\Delta \phi_{12}^{*}=1/2$, $\Delta \phi_{13}^{*}=3/4$ and $\Delta \phi_{14}^{*}=1/4$ for specific $g^{\rm inh}_{12}=g^{\rm inh}_{21}=2.25g_{\rm max}$, $g^{\rm elec}=0.25g_{\rm max}$, and nominal conductances $g=g_{\rm max}=2.5 \times 10^{-3}$.
}
\label{fig3}
\end{figure*}

We define a sequence of {\it phase-lags} through the delays in burst initiations relative to that of the reference neuron~1, normalized over the current network period, or, specifically, the burst recurrent times for the reference interneuron, as follows:
\begin{equation}
\Delta \phi^{(n)} _{1j} = \displaystyle{\frac{t^{(n+1)}_{j} - t^{(n)}_{1}}{t^{(n+1)}_{1} -t^{(n)} _{1}}}
\quad \mbox{mod 1,} \quad \mbox{where} \quad j=2,3,4.
\label{eq_lag}
\end{equation}
An ordered triple, ${\mathbf M}_{n} = \left (\Delta \phi^ {(n)} _{12}, \Delta \phi^{(n)} _{13}, \Delta \phi^{(n)} _{14} \right ) $, defines a forward iterate, or a phase point (see Fig.~\ref{sfig3b}), of the Poincar\'e return map for the phase-lags: ${\mathbf M}_n  \to {\mathbf M}_{n+1}$. A sequence, $\left \{(\Delta \phi^{(n)} _{12}, \Delta \phi^{(n)} _{13}  \Delta \phi^{(n)} _{14}) \right \}_{n=0}^N$, yields a forward
\emph{phase-lag trajectory}, $\left \{ \mathbf M_{n} \right \}_{n=0}^N$, of the Poincar\'e return map on a 3D torus  $[0,1)\times [0,1)\times [0,1)$ with phases defined on modulo~1, see Fig.~\ref{sfig4a}.

Based on the experimental recordings of interneurons of the {\it Melibe} swim CPG the authors \cite{SK11} suggest a possible architecture for its network. Its  wiring schematics shown in \ref{sfig3a}  includes two core half-center oscillators as the network building blocks: HCO1 (top, shown in blue) and HCO2 (bottom, shown in pink). The interneurons of each HCO burst robustly in anti-phase while the animal swims. The interneurons within a HCO are known to inhibit contralaterally each other: $1\,\bullet$---$\bullet\,2$ and $3\,\bullet$---$\bullet\,4$,  while 1 and 2 excite ---$\lhd$ 3 and 4 ipsilaterally, resp. In addition there is an electrical coupling through a gap junction between the top interneurons, 1 and 2. Strong electrical ipsilateral
coupling between 1 and 3, as well as 2 and 4 on the other side, makes them oscillate together. 

Traces of the bursting membrane potentials of the interneurons of the CPG are used to derive the phase-lags according 
to the method illustrated in  Fig.~\ref{fig1}. By varying the delays between burst initiation in the reference interneurons and release times of three other ones from inhibition, we obtain a dense array of initial phase distributions.  Then, the tuples of three  phase-lags are recorded at every cycle of the network bursting as the time progresses. Figure~\ref{sfig3c} illustrates a typical evolution of the resulting sequences,  
$\{\Delta \phi^{(n)}_{12}\}$,  $\{\Delta \phi^{(n)}_{13}\}$ and $\{\Delta \phi^{(n)}_{14}\}$ plotted against the burst cycle number, $n$, converging to a phase-locked state around $(1/2, 3/4,\,1/4)$; (unless otherwise mentioned, green, black and blue are the color-codes for the phase-lags, respectively.)

We  represent the sequence $\left \{ \mathbf M_{n} \right \}_{n=0}^N$ as a forward trajectory on a 3D-torus as the burst cycle number, $n$, is increased. Geometrically,  the 3D is viewed as a solid unit cube
shown in Fig.~\ref{sfig3b}. The opposite sides must be identified due to the cyclic nature of the phase; this implies that trajectory leaving the cube
through one of its six sides, will be wrapped around to re-enter through its opposite side and so forth.  

Alternatively, we can consider progressions of the phase lags individually, in terms of   
1D Poincar\'e return maps:  $\Delta \phi^{(n)}_{1j} \to \Delta \phi^{(n+1)}_{1j}$  (Fig.\ref{sfig3d}). Due to weak coupling,  sequential iterates $\{ \Delta \phi^{(n)}_{1j} \}$ of the maps do not jump far apart from each other. This allows for having them connected into ``continuous"  trajectories, in order to follow their evolution in forward time. The green sphere, in Fig.~\ref{sfig3b},
representing the initial phase-lags tuple, corresponds to the beginning of a trajectory.  
Such initial phase-lags are uniformly distributed on a lattice within a unit cube,  see Fig.~\ref{sfig3b}).   

The tuple, $\mathbf M_{n}$,  of the phase-lags represents the state of the network at the $n$-th burst cycle, because it captures the temporal bursting activity of all four neurons of the CPG. In what follows, we will explore visually how the network state progresses by following the evolutions of various initial phase-lags, which can converge to a single or multiple attractors.  Such an attractor of the map corresponds to a stable bursting pattern configurations of the CPGs in question. 

The initial distribution of network states or phase-lag tuples, sampled uniformly to form the cubic lattice within the cube, will shift toward attractors as the burst number increased. In the case when such an attractor is a fixed point, its coordinate corresponds to stable   phase-locked state of the bursting pattern with specific time delays between the interneurons of the CPG. It is equivalent that a single stable fixed point of the return map describes a single robust pattern of the dedicated CPG, as all initial delays will ultimately lead to the same bursting pattern. As the parameters of the CPG are changed, the stable fixed point can bifurcate, for example vanish or loose the stability, thus  giving rise to another attractor such as an invariant circle.  In the latter case, the phase of the bursting pattern are no longer locked,  but vary periodically, or even show some aperiodic, chaotic dynamics.  

If the corresponding return map shows two (or more) attractors, then depending on initial phases the multifunctional CPG can respectively produce several patterns. Perturbations, such as noise or external polarized currents, can cause sudden or unforeseen jumps between the attractors, resulting in 
switching between the corresponding  patterns, for example between in-phase and out-of-phase bursting.

It is worth noticing that in weakly coupled cases, one should consider constructing 1D Poincar\'e maps for every $k$-th phase-lags. This procedure gives a map of the degree $k$, with a ``flatter," so to speak, graph adjoining the $45^\circ$-line  at a stable fixed point.  Figure~\ref{sfig3d} displays such maps of degrees 1, 5 and 10, with the corresponding fixed points around $1/2$, $3/4$ and $1/4$  for the individual phase-lags as predicted from trace progressions depicted in Fig.~\ref{sfig3c}. Note that the fixed points are  approached from one side only. 
 
It is our working hypothesis that a stable state of the CPG is defined by interplays of synaptic strengths, rather than by specific wiring of the network. While wiring can be a necessary condition for the network to produce bursting pattern(s) , its configuration does not provide the sufficient condition for the robustness of the latter.  Thus, the problem of the robustness of the pattern  is reduced to the stability conditions and identification of bifurcations of the corresponding
fixed points of the map for the phase-lags. The current state is such that the maps have to be visualized to identify and classify 
all patterns resulting from different initial phase relations between the four bursting interneurons.

\begin{figure*}[!htbp]
  \centering
   {\resizebox*{.6\columnwidth}{!}{\includegraphics{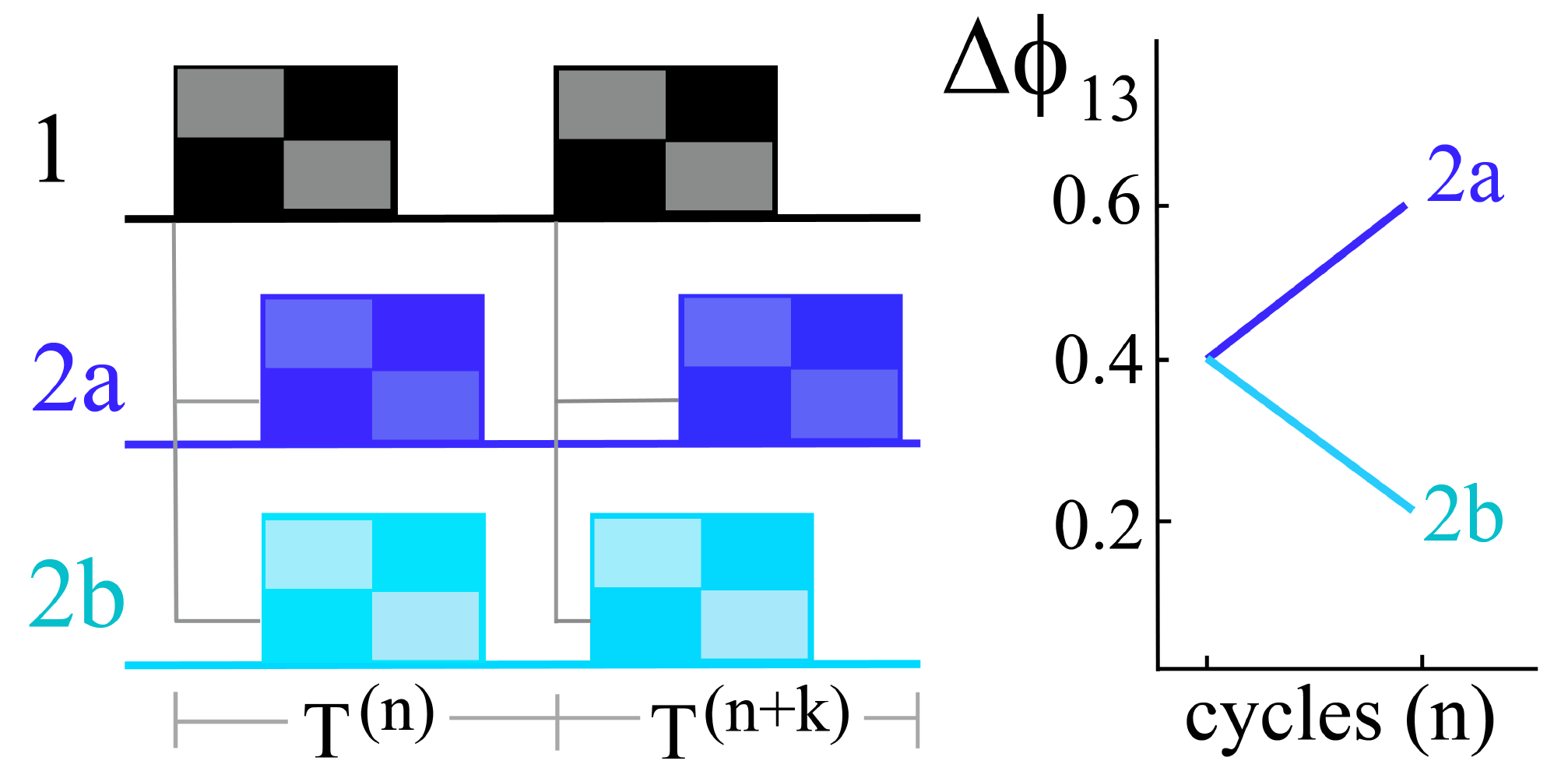}}}
  \caption{(left) Evolutions of the phase-lags between the HCOs  depending on the ratio of their periods, plotted against each $k$-th 
cycle. Checkered rectangles, black/gray and blue/light blue represent time progressions of relative anti-phase bursting in the HCOs when 
the period of HCO2 (cyan/blue) is initially longer (2a), or shorter (2b)  than the period of HCO1 (black). (right) Sketch for
increasing/decreasing change in the value of the phase lag, $\Delta \phi_{13}^{(n)}$, depending on the ratio of the periods of HCOs.}
  \label{fig2}
\end{figure*}

The following concern must be also  addressed: how can changes in phase-lags relate to the periods of the two HCOs forming the 4-neuron CPG. Namely, whether it is crucial that the periods of both individual HCOs are not the same, and how coupling affects the period of the whole network.  Figure~\ref{fig2} addresses this issue: let HCO2 have the period, $T_2$, slightly longer (shorter) than the period, $T_1$ of HCO1 (i.e. $T_1/ T_2>1$ or $T_1/ T_2<1$), then phase-lags between them, here $\Delta \phi_{13}^{(n)}$, will increase (decrease) in the CPG, which begins 
from the same initial conditions.   This relation between the phase-lags and the periods  is discussed in Appendix.

\section{Results}

The main question that we aim to address in our modeling study is: what synaptic connections are the key ones that lead to the experimentally observed (i.e. stable) phase-locked bursting in the CPG. 

Let us first consider a network configuration of the CPG with only contralateral inhibitory 
connections of the strength $g^{\rm inh}_{32}=g^{\rm inh}_{41}=2.5\times10^{-3}$, which is a half of that of the 
strongest inhibitory synapses in the CPG ($g^{\rm inh}_{34}=g^{\rm inh}_{43}=5\times10^{-3}$).  
Later, we introduce the ipsilateral excitatory synapses ($g^{\rm exc}_{13}$ and $g^{\rm exc}_{24}$), followed by the electrical synapses or gap junction ($g^{\rm elec}$) between the interneurons 1 and 2 to examine transformations of bursting patterns through  
bifurcations,  if any, of attractors in the corresponding 3D maps on the torus.    

\begin{figure*}[!htbp]
  \centering
  \subfigure [][] {\resizebox*{.45\columnwidth }{!}{\includegraphics{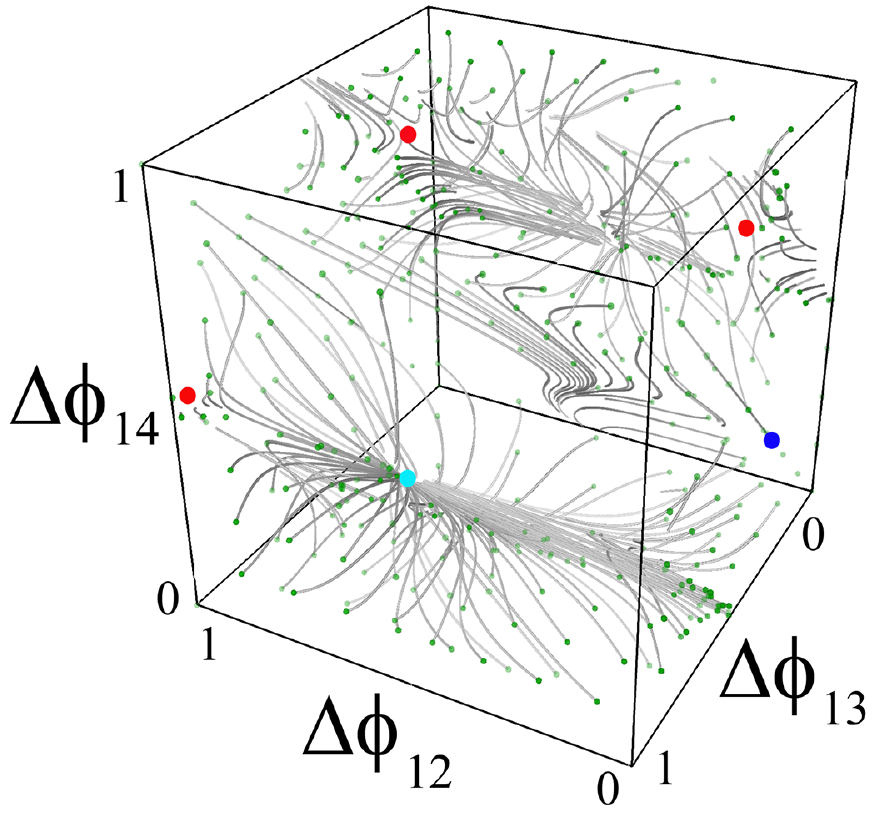}} \label{sfig4a}}~~~~~
%  \subfigure [][] {\resizebox*{.43\columnwidth}{!}{\includegraphics{fig4bn.pdf}} \label{sfig4n}} 
  \caption{(a) Phase space of the return map for the inhibitory CPG in the heterogenous case with balanced inhibition. Trajectories leaving one side of the cube (torus) are wrapped around to re-enter from the opposite side.  Red dots indicate the locations of turning points -- saddles,  the cyan one locates the global attractor at $ (\Delta \phi_{12},\,\Delta \phi_{13},\,\Delta \phi_{14})= (1/2,\,0,1/2)$  corresponding to the phase-locked state of the bursting  pattern as such shown in (b). The blue dot is an attractor at $(0,1/4,1/4)$ with a narrow basin corresponding to 4-cell CPG 
with the in-phase HCO driven in anti-phase by HCO bursting in-phase too. Here, the conductances are $g^{\rm inh}_{12}=g^{\rm inh}_{21}=2.5\times10^{-3}$, $g^{\rm inh}_{32}=g^{\rm inh}_{41}=2.5\times10^{-3}$ and $g^{\rm inh}_{34}=g^{\rm inh}_{43}=5\times10^{-3}$. (b) (a) Synchronous bursting pattern intracellularly recorded from the identified swim interneurons Si1-2LR of the Melibe CPG \cite{NSLGK012} (with the locked phase-lags corresponding to the coordinates of the fixed point attractor of the 3D map in (a)). Recording provided courtesy of A.~Sakurai. }
  \label{fig4}
\end{figure*}

We have performed comprehensive simulations and further visualization of solutions of return maps for the sequences of phase-lags for the   CPG configurations. For the given CPG model, the corresponding 3D Poincar\'e map is shown in Fig.~\ref{sfig4a}. It displays multiple transients 
converging to a stable fixed point corresponding to the phase-locked lags within the bursting pattern. In this figure, a green dot indicates the beginning point of a transient of the map in every simulation run. By releasing trajectories from a dense, homogeneously distributed grid
of initial phases (conditions) spread over the bursting periodic orbits, one can obtain a complete portrait of the phase space of the return map
on the unit cube. This 3D portrait in Fig.~\ref{sfig4a} shows attractors, separating saddles and invariant subspaces. Here,  the red  circles indicate the locations of saddles, or turning points, while the cyan and the blue circles correspond to the steady states of the network, i.e. stable fixed points of the map. 

The fixed point represented by the solid cyan circle in Fig.~\ref{sfig4a} has the following coordinates: $\Delta \phi_{12} = \Delta \phi_{14} =1/2$
and $\Delta \phi_{13} = 1 \equiv 0$.  This means that the interneuron~1 and 3 burst in-phase with respect to each other, while they keep bursting in anti-phase with their counterparts: interneurons 2 and 4. In other words, the CPG is made of the HCOs bursting in-phase, synchronously, 
with respect to each other.   By inspecting the stability of the fixed point in restriction to the side,  $\Delta \phi_{13} = 1$ of the cube, one can conclude that  there are two ways which give rise to this bursting rhythm. The independent HCO2 can either slow down, or catch up, to line up with the inhibitorily driven HCO1. One can  conclude  here that, in essence, the phase relations between the interneurons of this CPG is effectively reduced to that between the HCOs, provided that both  maintain anti-phase bursting endogenously.     
 
There is another stable fixed point of a smaller basin in the phase space of the map for this CPG configuration. Its location is indicated by the solid blue circle, $\Delta \phi_{13} = \Delta \phi_{14}=1/4$, of the side, given by $\Delta \phi_{12} = 0 \equiv 1$, of the cube. The coordinates of this fixed point correspond to a rather constrained rhythm:  the in-phase bursting interneurons 1 and 2 of HCO1 are driven, causing a $1/4$ phase-lag, by the also in-phase bursting interneurons 3 and 4 of HCO2. Therefore, near this fixed point the 4-cell CPG acts as a 2-cell network, which is equivalent to one interneuron inhibiting the other with a double drive.       

Some projections of the map on the unit cube can be misleading for evaluations of the locations of the stable fixed points, as one has to see the orthogonal projections, along with the phase-lag progressions plotted against the bursting cycle. Note that interpretations of such progressions
(projections too) could be a challenge in a multi-stable case where overlapping trajectories tend to converge to several attractors. To give a comprehensive overview of the dynamics of the 3D map we will also  utilize such orthogonal 2D projections and frequency coint distributions to understand better the behavior of its transients. Of particular interest are how they converge to the attractors, as in some cases the convergence can be achieved only from one side. In terms of the map for a dedicated CPG, this means strengthening the stability of the single fixed point from any direction in the unit cube, which becomes its global attraction basin. Addition of synapses to the CPG may make it multifunctional, so it is 
imperative to know how this changes quantitatively the number of attractors and qualitatively the stability conditions for the bursting patterns.
 
We start the next section with a discussion on a dissected CPG with uncoupled HCOs. 
While this may seem trivial, it should give us a reference framework necessary for singling out the underlying organizations of the phase-lag trajectories resulting from addition of basic synaptic connections. In the 4-cell network, the connections, contralateral inhibitory and ipsilateral excitatory, should promote, not conflict, the robustness of bursting  outcomes of the CPG.

\subsection{CPG dissection in homo- and heterogeneous HCOs}

A network state will correspond to a behavior pattern if it evolves on par with the behavior. By introducing variations in carefully chosen synaptic connections of the network, we make predictions and match outcomes with the expected behaviors, in order to identify the CPG mechanism. In this study, we assume that a persistent phase-locked state underlies the {\it Melibe} swimming behavior. While the ideal mathematical model must reflect all experimentally observed features of the biological CPG, a reduced model is intended to describe only some likely mechanisms  giving rise to  stable phase-locked bursting patterns such as the one depicted in Fig.~\ref{fig1}.

In order to elucidate how CPG networks operate in general,  and, in particular, how  the {\it Melibe} CPG, robustly produces the single pattern with the constant  phase-lag, we apply a {\em bottom-up} approach. This approach is used for identifying and differentiating the features that persist as the network configuration becomes more plausible in comparison with the biological CPG architecture. For example, one pair of uncoupled HCOs suffices to produce anti-phase bursting patterns observed in the voltage traces recorded from four interneurons. In addition, the capacity  of pattern generation of 3-neuron  motifs made of reciprocally inhibitory interneurons is well understood \cite{SGB08,WCS11}. 
 
It was shown recently that under certain  conditions, fast non-delayed reciprocal inhibition within a stand-alone pair of similar neurons may lead to synchronous, in-phase bursting \cite{pre2010,pre2012}. So, for the sake of generality, we set the parameters of the individual interneurons and the cross-coupling some different to guarantee that anti-phase bursting is the only stable pattern in either HCO  \cite{BS08, WCS11}.
 
\begin{figure*}[!htbp]
  \centering
  \subfigure [][] {\resizebox*{.2\columnwidth}{!} {\includegraphics{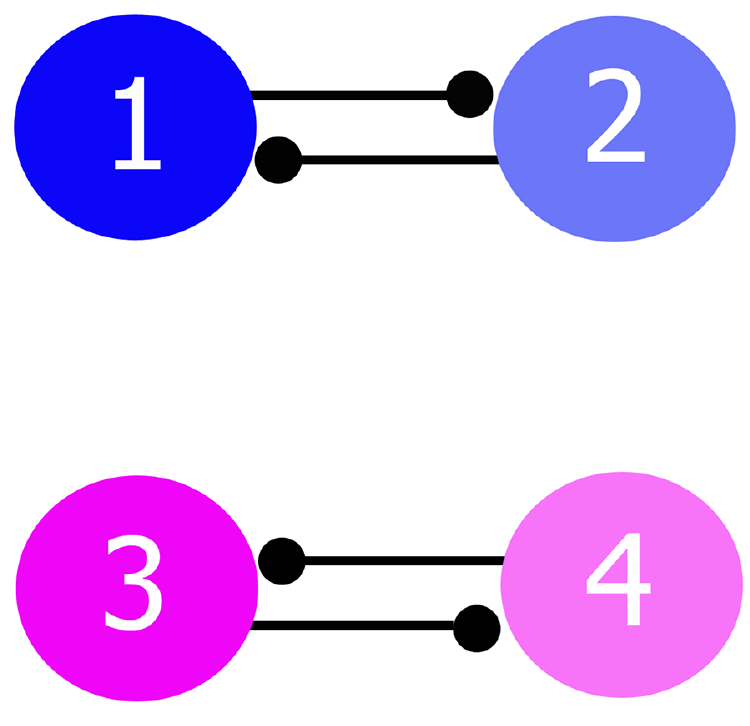}} \label{sfig5}}
  \subfigure [][] {\resizebox*{.5\columnwidth}{!} {\includegraphics{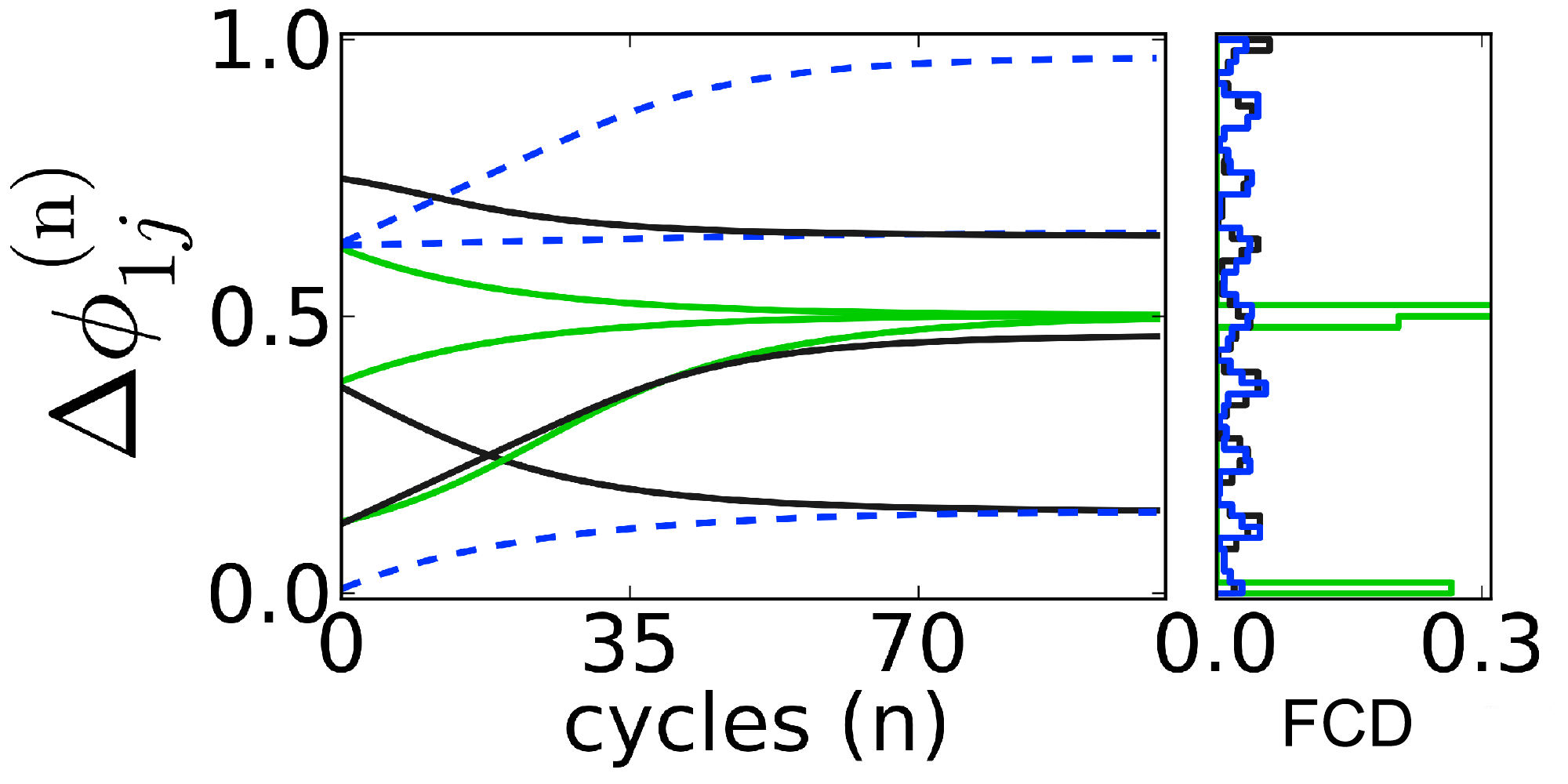}} \label{sfig8}}\\
  \subfigure [][] {\resizebox*{.45\columnwidth}{!}{\includegraphics{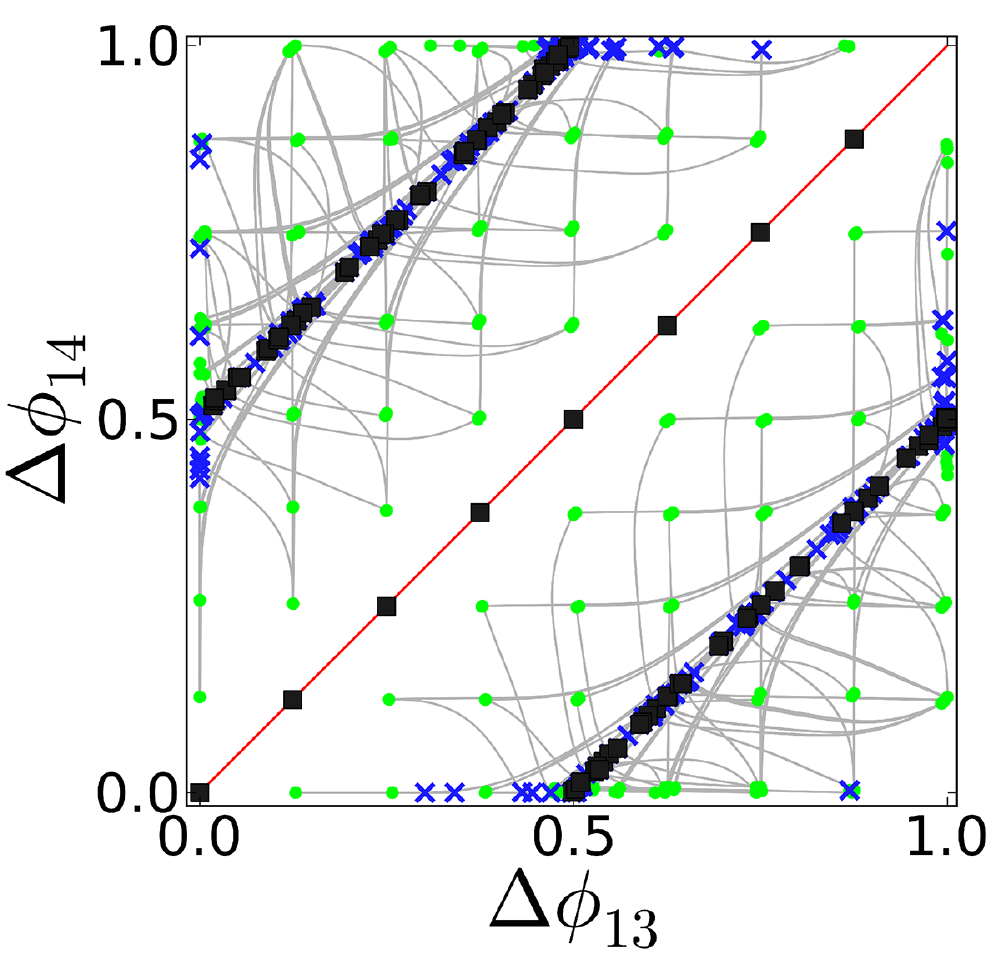}} \label{sfig6}}  
  \subfigure [][] {\resizebox*{.4\columnwidth}{!} {\includegraphics{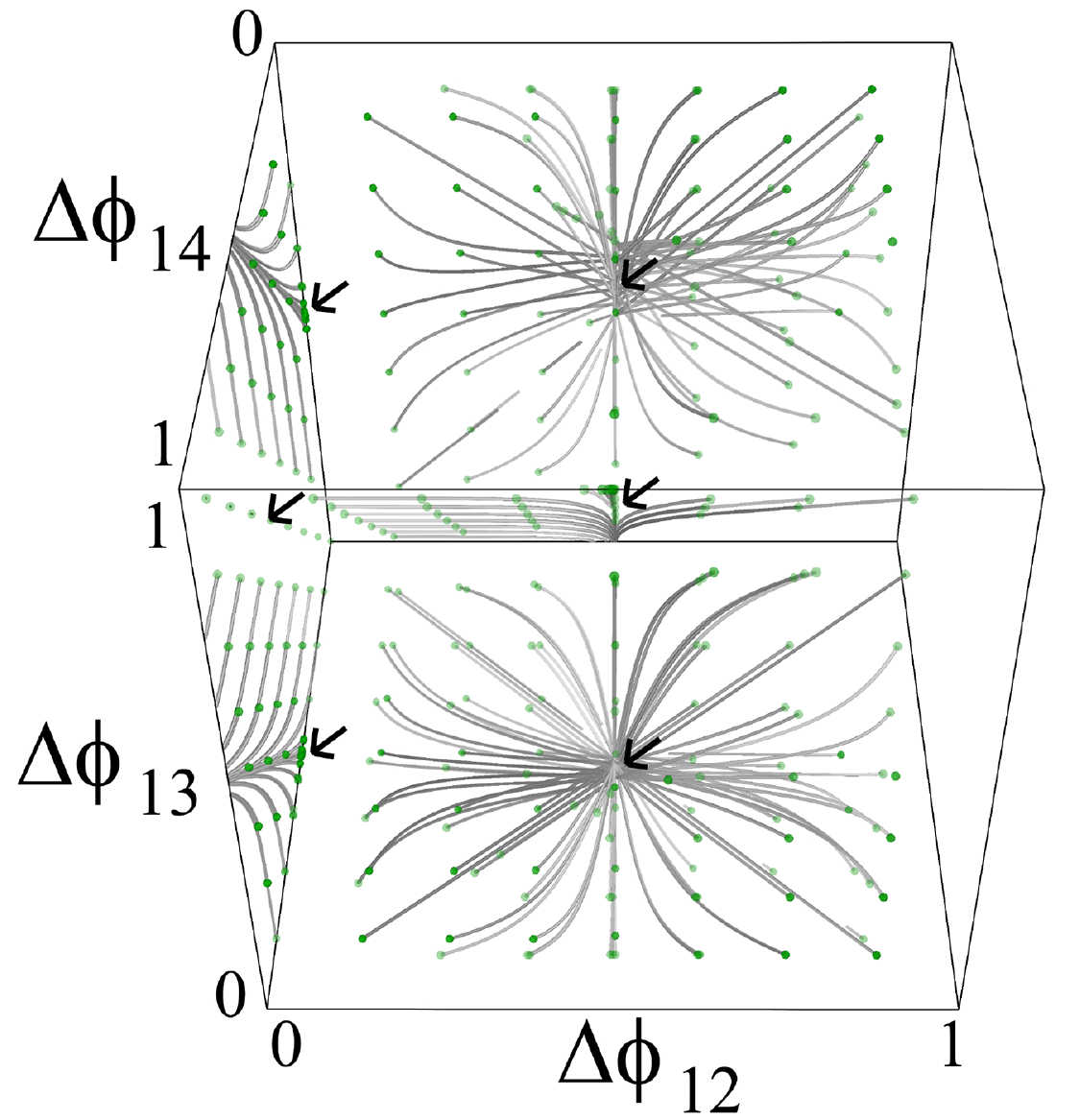}} \label{sfig7}}
  \caption{(a) Dissected CPG made of uncoupled HCOs: $g^{\rm inh}_{13}=g^{\rm inh}_{24}=0$. (b) Phase-lag progressions (green, black and blue curves for $\Delta \phi_{12}$, $\Delta \phi_{13},$ and $\Delta \phi_{14})$, resp.) indicating multiple phase-locked states, and the frequency count distribution (FCD) of terminal phase-lags with two distinctive green peaks corresponding to in- and anti-phase  states in HCO1  for  $g^{\rm inh}=5\times~10^{-4}\,(1+\delta_{ij})$, $g_{\rm max}=5\times~10^{-4}$.  (c) 2D $(\Delta \phi_{13},\, \Delta \phi_{14})$-projection of the phase space of the map: green dots unmask the lattice of initial phases; blue crosses indicate the terminal points of (40 burst cycle long) trajectories on the line $\Delta \phi_{14}=\Delta \phi_{13} + 1/2$ (mod~1), corresponding to anti-phase bursting HCO2, and the (red) bisectrix corresponds to the in-phase bursting  HCO2. Black squares  indicate stagnation areas. (d) 3D return map revealing an attractor with $\Delta \phi_{12}=0.5$ corresponding to anti-phase bursting in HCO1.}
  \label{fig5}
\end{figure*}

Figure~\ref{fig5} recaps some findings for the dissected CPG made of uncoupled ($g_{32}^{\rm inh}=g_{41}^{\rm inh}=0$) and homogeneous  HCOs: all
maximal conductances are equal $g^{\rm inh}_{max}=5\times~10^{-4}$. Figure~\ref{sfig8} shows a few samples of the phase-lags progression,  $\Delta \phi_{12}$, $\Delta \phi_{13}$ and $\Delta \phi_{14}$, plotted against the cycle number. It shows that the network transients converge to more than a single attractor. 
Observe too that convergence rates to the phase-locked states are predictably equal in this homogeneous case.  Next to it is the frequency count distribution of 448 initial network states after 100 burst cycles. The diagram depicts two dominating peaks in green for $\Delta \phi_{12}$ ($\Delta \phi_{34}$) corresponding, respectively, to in-phase  and anti-phase bursting in HCO1 (HCO2). As the HCOs are uncoupled, the distributions for $\Delta \phi_{13}$ (in black) and $\Delta \phi_{14}$ (in blue) are uniform.  Since the HCOs receive no inputs from each other, they evolve independently resulting in the phase-lags between the reference neurons in each to be arbitrary.

 Figure~\ref{sfig6} represents the 2D $(\Delta \phi_{13},\, \Delta \phi_{14})$-projection
of the phase-lag transients. In it,  green dots unmask a uniformly distributed lattice of initial states of the network, 
 while blue crosses mark the ends of the phase-lag transients. Note again that when, a phase-lag goes over 1, its value is reset by modulo 1. Black squares in Fig.~\ref{sfig6} indicate regions of rather slow evolution of the network transients, where sum of the phase-lags per cycle shift less than $0.005$ over the last ten cycles.

The 3D phase space of the unit cube for the uncoupled HCOs is given in Fig.~\ref{sfig7}. It shows two nodal attractors 
corresponding to the anti-phase bursting in each HCO. Because of the equal coupling weights, each node is ``symmetric" thus indicating even convergence rates in both HCOs.  The lack of interaction between HCO explains the presence of an invariant curve, given by the constraints $\Delta \phi_{14}=\Delta \phi_{13} + 1/2$ (mod~1) and $\Delta \phi_{12}=1/2$. This invariant line corresponds to anti-phase bursting in the uncoupled HCOs. The curve is stable in  directions transverse to it, and neutrally stable along it, i.e. the phase-lags on it do not shift in the homogeneous case.  In other words, each HCO tends typically, for most initial conditions, to anti-phase bursting. 

\begin{figure*}[!htbp]
\centering
\subfigure [][] {\resizebox*{.45\columnwidth}{!}{\includegraphics{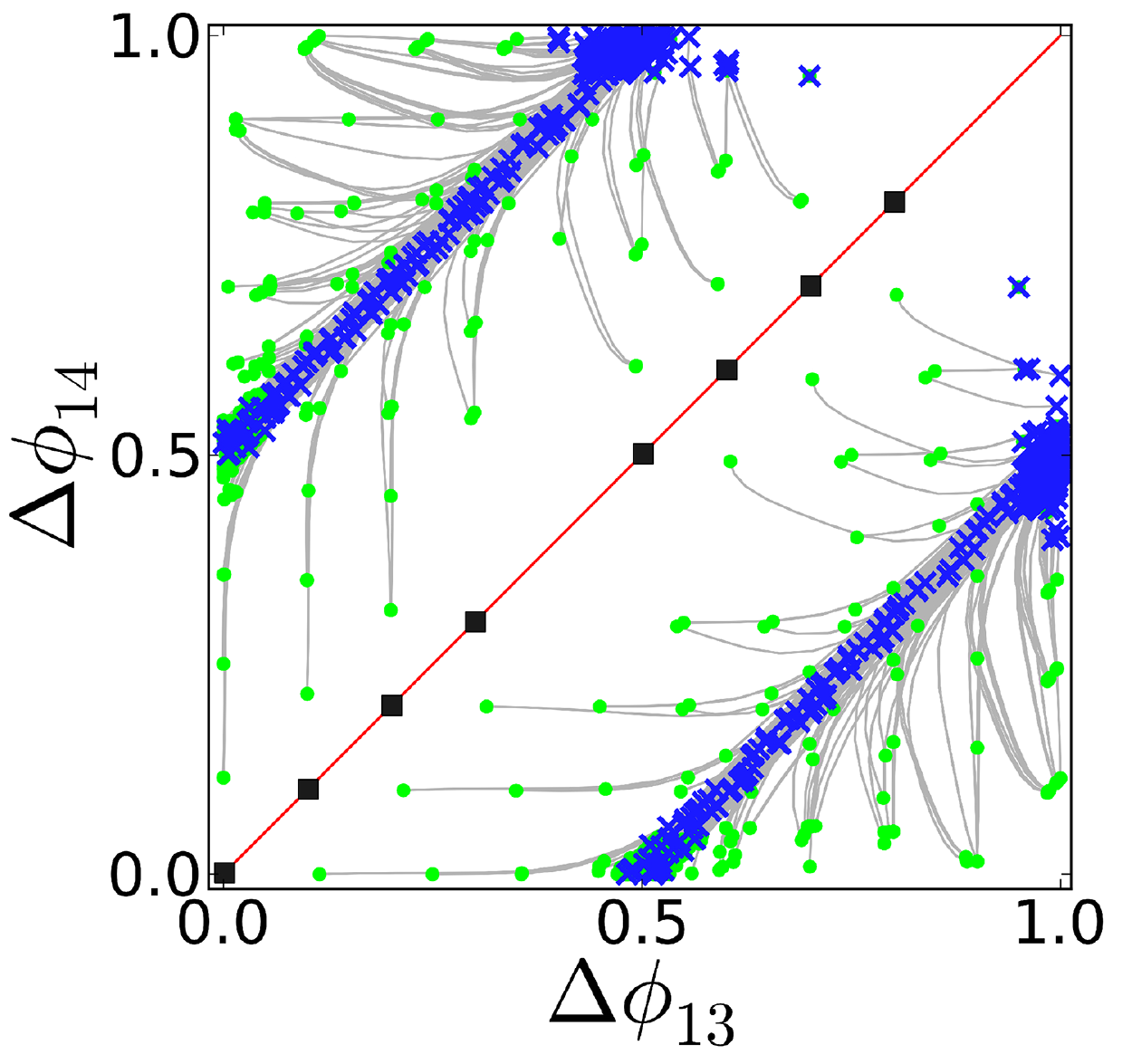}} \label{sfig7c}}
\subfigure [][] {\resizebox*{.45\columnwidth}{!}{\includegraphics{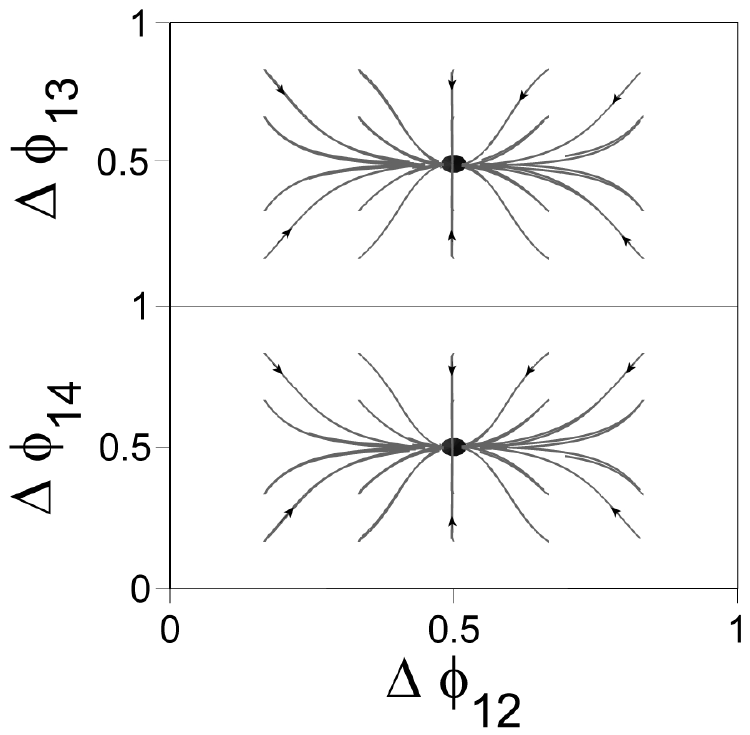}} \label{sfig7d}}
\caption{(a) $(\Delta \phi_{13},\, \Delta \phi_{14})$ phase-lag projection for a heterogeneous CPG at $g_{12}^{\rm inh}=g_{21}^{\rm inh}=0.5g_{\rm max}$.
 Blue dots indicate the terminal states of transients (gray lines) originating from initial  (green) states. Having converged to the line   $\Delta \phi_{14}=\Delta \phi_{13} + 1/2$ (mod~1), (40 burst cycle long) transients slide slowly along it. (b) 2D  $(\Delta \phi_{12},\, \Delta \phi_{13})$ and  $(\Delta \phi_{12},\, \Delta \phi_{14})$ projections showing the stable fixed point at (1/2,\,1/2) with the dominant (horizontal) leading and strongly stable (vertical) directions due to the distinct coupling strengths in the HCOs, i.e. the convergence rate to the anti-phase bursting in HCO2 with $g_{\rm max}$ is faster than that in HCO1 with $0.5\,g_{\rm max}$.}
  \label{fig7}
\end{figure*}

Next, let us consider a heterogeneous CPG such that the reciprocal inhibitions in HCO1 are two times less than those in HCO2:   
$g_{\rm HCO1}^{\rm inh}=0.5\,g_{\rm HCO2}^{\rm inh}=g_{\rm max}(1+\delta_{ij})$, where $g_{\rm max}=5\times 10^{-4}$. The corresponding return mappings are shown in
Fig.~\ref{fig7}. Symmetric $(\Delta \phi_{12},\, \Delta \phi_{13})$ and $(\Delta \phi_{12},\, \Delta \phi_{14})$ phase-lag projection shows the persistent
nodal attractor corresponding to the endogenously anti-phase bursting HCOs. The quantitative changes of the heterogeneous case compared to the outcomes of homogeneous one 
is that the fixed point has  the leading (horizontal) and strongly stable (vertical) directions (Fig.~\ref{sfig7d}) due to, correspondingly, faster and slow convergence rates to the anti-phase bursting in HCO2 and HCO1 with stronger and weaker reciprocally inhibitory synapses.  As a result, transients, which have converged to the invariant line, 
 $\Delta \phi_{14}=\Delta \phi_{13} + 1/2$ (mod~1)  (Fig.~\ref{sfig7c}) slide slowly along it. The slow speed rate is proportional to the ratio
 of the bursting periods of the uncoupled HCOs, which are no longer equal in the heterogeneous case.   

 \section{Coupled Inhibitory CPGs}

Restoring the contralateral inhibitory connections, $g^{\rm inh}_{32}=g^{\rm inh}_{41}=g_{\rm max}(1+\delta_{ij})$, feed-forwarded from the driving HCO2 to the driven HCO1  enhances  the robustness of the bursting patterns of such CPG. Depending on whether it is comprised of homo- or heterogeneous HCOs, the phase-lags between the HCOs can vary, thus giving rise to principally distinct patterns.  The unidirectional inhibition gives rise to a polarity in the network.  As a result, burst timing of the driven HCO1  has to adjusts itself to that of the driving HCO2  in order for the network to settle into a steady rhythm, if any,  with all phases locked.  Contralaterality of such inhibition is significant due to the effect it has on timing of interactions when (balancing) ipsilateral  excitation is rewired in the CPG schematics as in Fig.~\ref{sfig3a}.

\begin{figure*}[!htbp]
  \centering
  \subfigure [][] {\resizebox*{.45\columnwidth}{!}{\includegraphics{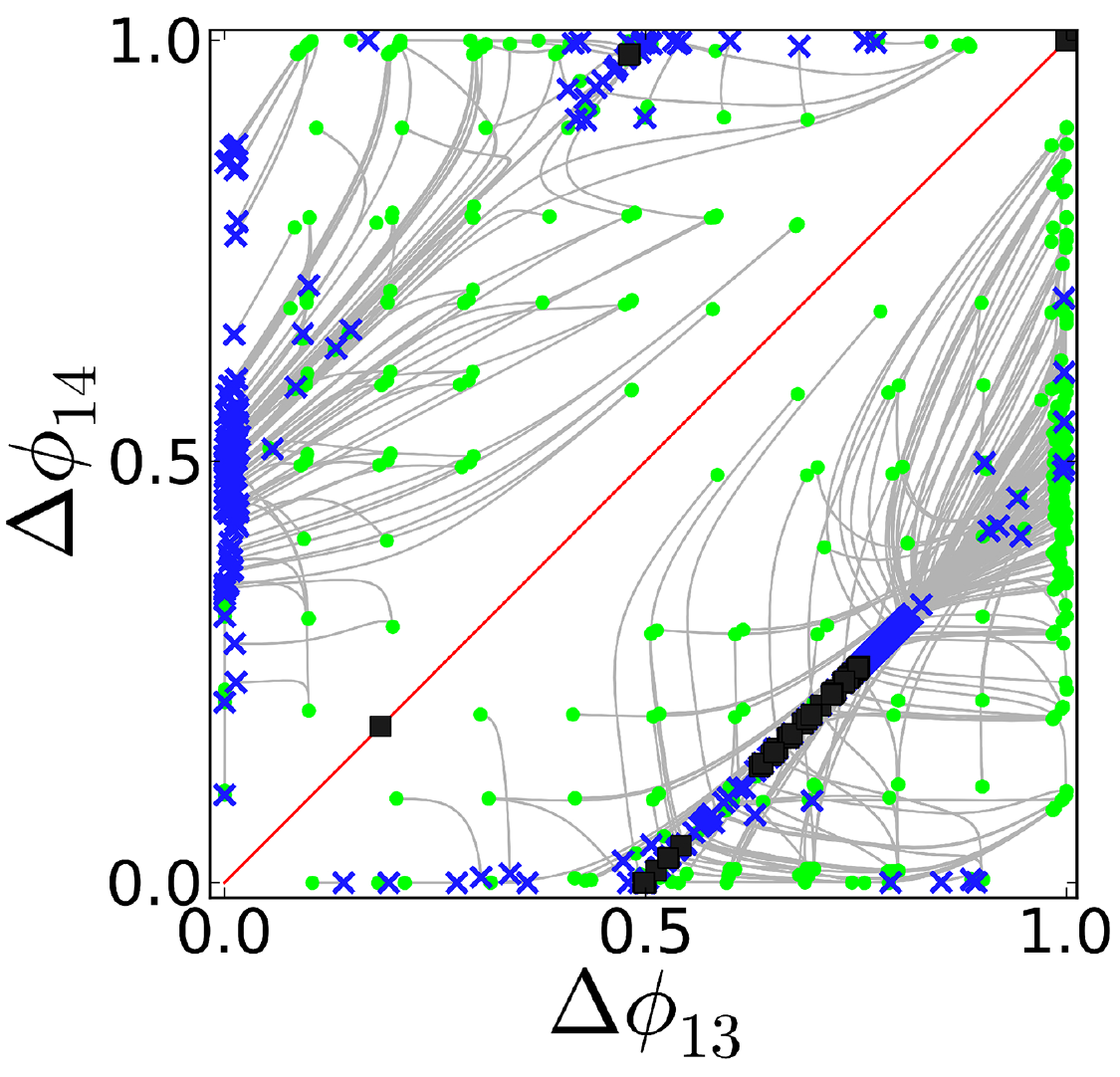}} \label{sfig11}}
  \subfigure [][] {\resizebox*{.5\columnwidth}{!}{\includegraphics {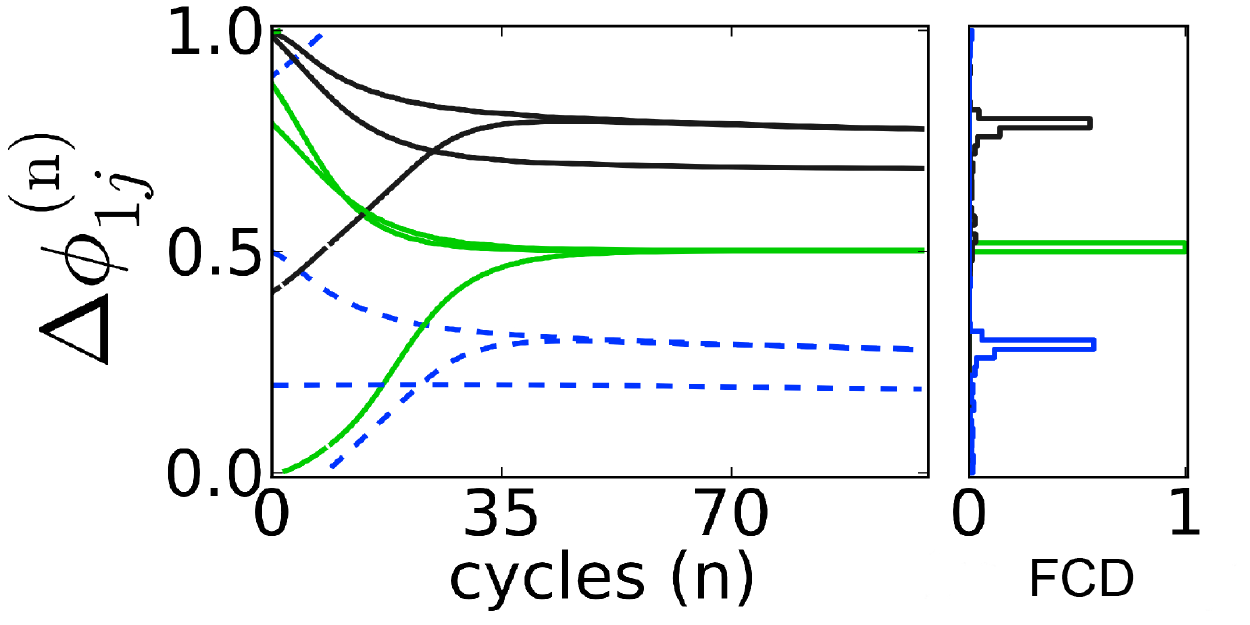}} \label{sfig12}}\\
  \subfigure [][] {\resizebox*{.45\columnwidth}{!}{\includegraphics{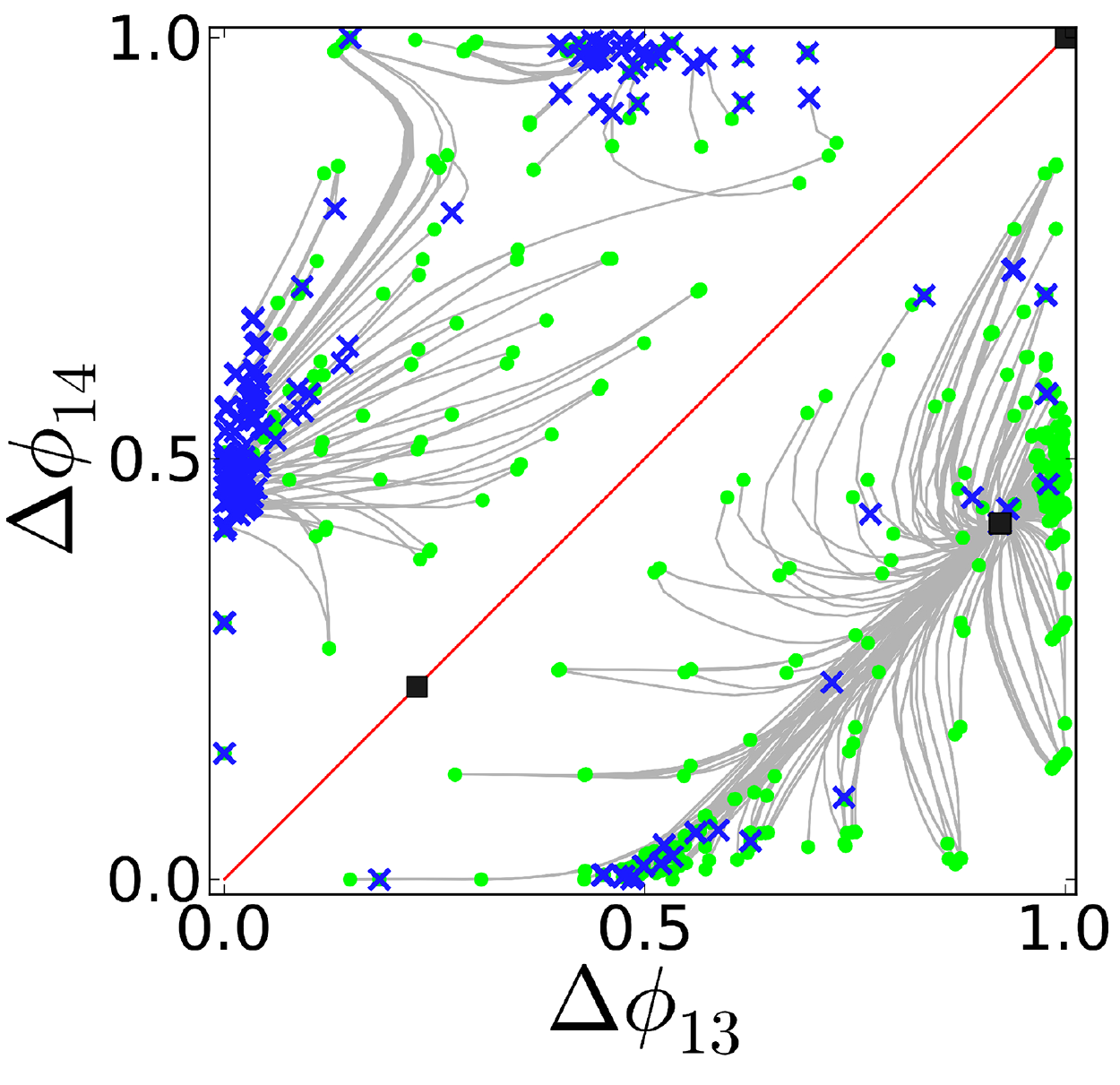}} \label{sfig15}}
  \subfigure [][] {\resizebox*{.5\columnwidth}{!}{\includegraphics {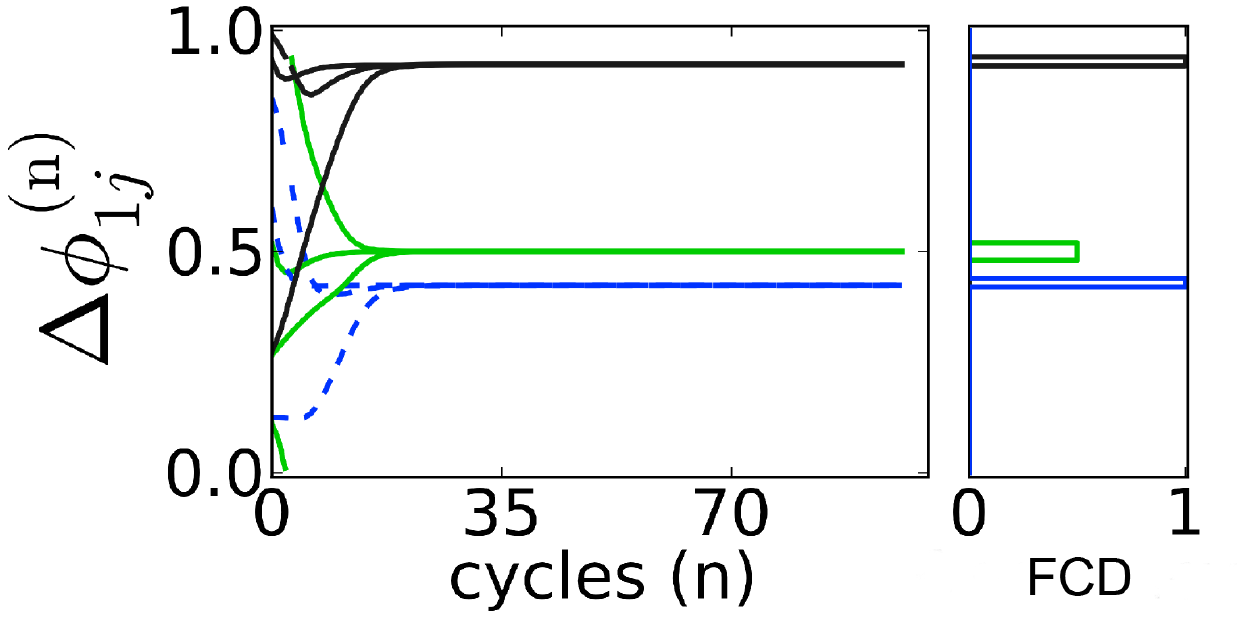}} \label{sfig16}}
  \caption{(a,c) $(\Delta \phi_{13},\, \Delta \phi_{14})$-projections of the phase-lag maps for the contralaterally inhibitory CPGs (circuitry digram shown in Fig.~\ref{sfig17a}) made of homo- and heterogeneous HCOs. Transients (grey lines), having bended around saddles, converge to the unique stable fixed point (blue crosses) located, respectively, at ($1/2,\,3/4,\,1/4$) and approximately at ($1/2,1 \equiv 0,\,1/2$).  (b,d)  Left panels: representative  samples of  phase-lag progressions (green, black and blue curves for $\Delta \phi_{12}$, $\Delta \phi_{13},$ and $\Delta \phi_{14})$, resp.) and the frequency count distribution (FCD)  (right panels) of the terminal states of phase-lags to identify the coordinates of the fixed points of the return maps.}  \label{fig8}
\end{figure*}

Figure~\ref{fig8} represents the $(\Delta \phi_{13},\, \Delta \phi_{14})$-projections of the phase-lag maps for homogeneous (a) and heterogeneous (c) networks.  
In the former case, the majority of the transients tend to a single fixed point at $(\Delta \phi_{12},\,\Delta \phi_{13},\,\Delta \phi_{14})=(1/2,\, 3/4,\,1/4)$. 
Still, there is a rudiment of the invariant curve segment, $\Delta \phi_{14}=\Delta \phi_{13} - 1/2$, due to equal convergence rates in the HCOs.
These values of the fixed point coordinates are supported  by inspection of a few delegated phase-leg progressions plotted against the burst cycle number. Figure~\ref{sfig12} yields  the frequency count distribution (FCD) of the network states after 100 cycles. The diagram shows a sharp peak at $\Delta \phi_{12}=1/2$, along with  wider 
peaks (black and blue) at $\Delta \phi_{13}=3/4$ and $\Delta \phi_{14}=1/4$.  The values of the coordinates of the fixed point mean that 
unidirectional inhibition shifts anti-phase bursting in the driven HCO1 a quarter of the period forward relative to that of the driving HCO2 in the homogeneous network with the same synaptic conductances.  

The heterogeneous CPG demonstrates other phase-lags between the  HCOs.  Recall that in this case the reciprocal inhibitions in HCO1 are halves of those in HCO2:    
$g_{12}^{\rm inh}=g_{21}^{\rm inh}=0.5g_{\rm max}(1+\delta_{ij})$.  The corresponding phase-lag map in the $(\Delta \phi_{13},\, \Delta \phi_{14})$-projection given in  Fig.~\ref{sfig15} shows  no indication of an invariant line but the occurrence of a stand-alone stable fixed point. The strong stability of the fixed point is also supported
by the fast convergence of transients  to the steady states  (Fig.~\ref{sfig16}) after about 10 burst cycles, in contrast to 35 in the homogeneous case.  The sharp peaks, at or near $1/2$ for $\Delta \phi_{12}$ and $\Delta \phi_{14}$, and  near $1$ for $\Delta \phi_{13}$ in the frequency count distribution of terminal phase-lag is the secondary backup for this assertion.  Stated another way, the coordinates, $(\Delta \phi_{12},\Delta \phi_{13},\Delta \phi_{14}) \approx (1/2,\, 0\equiv 1,\, 1/2)$, of this fixed point correspond to the CPG rhythm where both   HCOs burst in-phase with the common period.

Next, we will explore how the uni-directional contralateral inhibition and ipsilateral excitation can balance out the CPG dynamics when introduced separately. This should let us identify independent contributions of the synapses of each type to the behavior of the  whole CPG.

\section{Basic heterogeneous inhibitory and excitatory networks}

\begin{figure*}[!htbp]
\centering
\subfigure [][] {\resizebox*{.2\columnwidth}{!}{\includegraphics{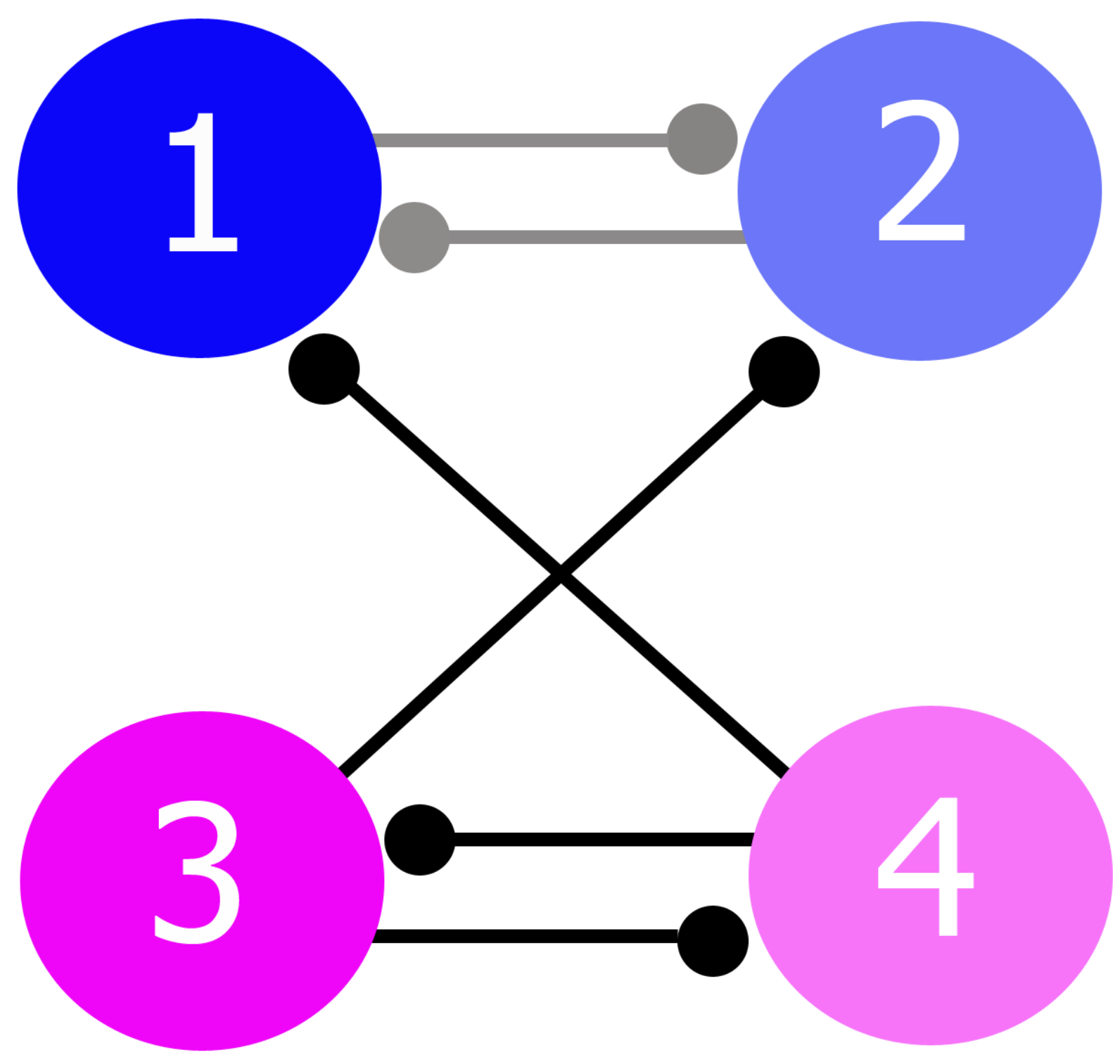}} \label{sfig17a}}
\subfigure [][] {\resizebox*{.5\columnwidth}{!}{\includegraphics{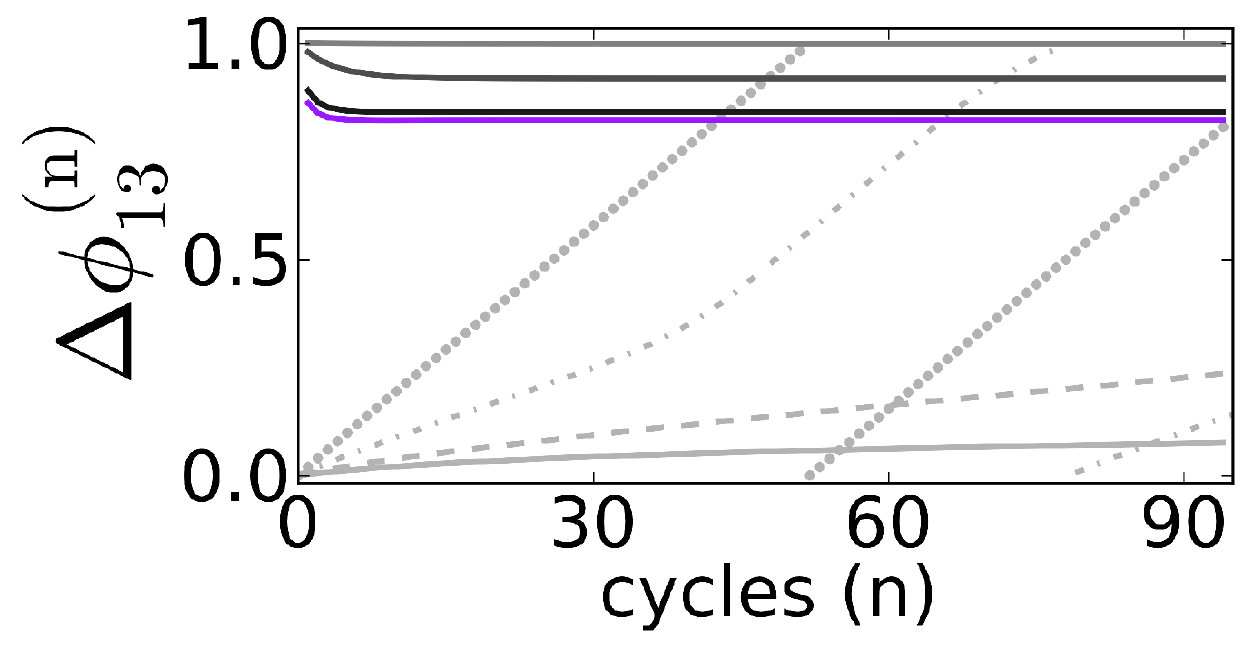}} \label{sfig17}}\\
\subfigure [][] {\resizebox*{.2\columnwidth}{!}{\includegraphics{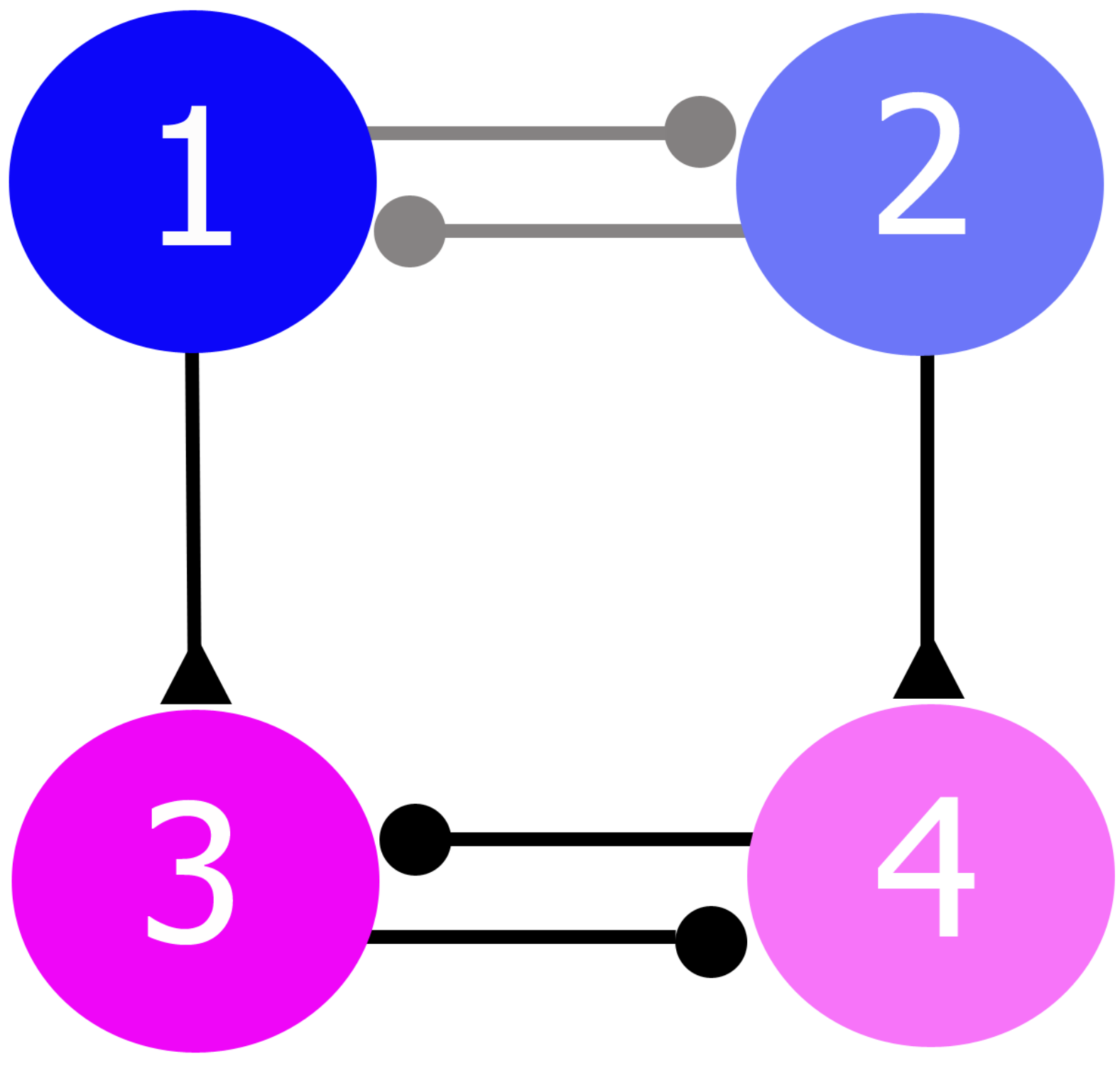}} \label{sfig18a}}
\subfigure [][] {\resizebox*{.5\columnwidth}{!}{\includegraphics{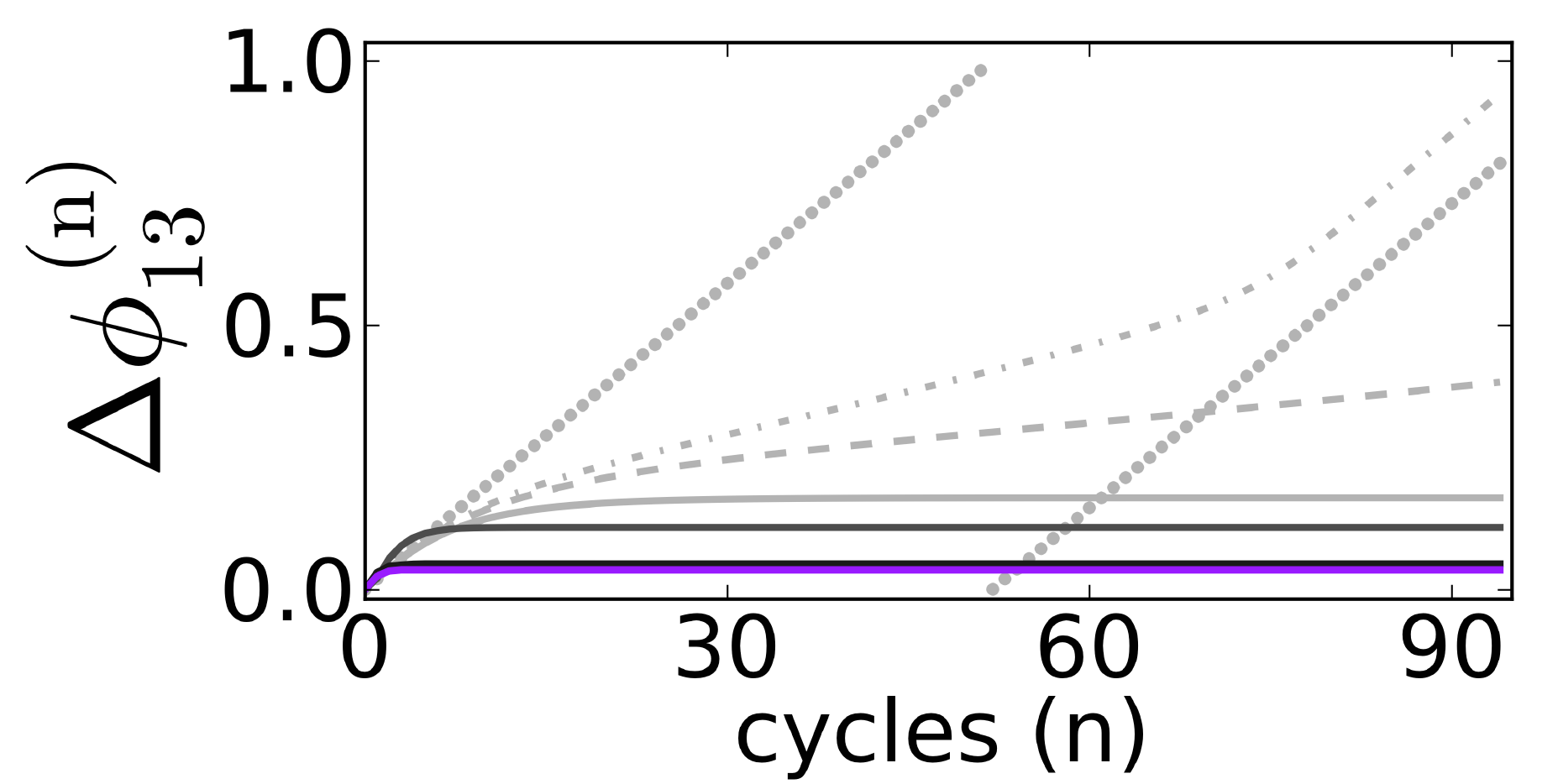}} \label{sfig18}}
\caption{ (a,c) Elementary circuitries of {\it Melibe} and {\it Dendranotus} CPG networks with feed-forward, contralateral inhibition 
for  and  reverse-feed ipsilateral excitatory synapses. 
(b,\,c)  Increasing contralateral inhibition strength, $g_{32}^{\rm inh}=g_{41}^{\rm inh}$, makes  HCO1 follow  HCO2,  while 
increasing ipsilateral excitatory synapses, $g_{13}^{\rm exc}=g_{24}^{\rm exc}$, reverses the order,  making the 
driven HCO2 follow the driving HCO1. (b) Evolutions of the phase-lag, $\Delta \phi_{13}$, converging to a high steady state as the contralateral 
inhibitory coupling is increased: $g_{32}^{\rm inh}=g_{41}^{\rm inh}=d \times g_{\rm max}(1+\delta_{ij})$,  
$d=0$ (grey dots), $0.25$ (grey dash-dots), $0.375$ (grey dashes), $0.425$ (solid grey), $0.5$ (dark grey),  $5$ (black), and $10$ (purple). (d) 
Evolutions of the  phase-lag, $\Delta \phi_{13}$, converging to a low steady state as the ipsilateral excitatory coupling is increased: 
$g_{13}^{\rm exc}=g_{24}^{\rm exc}=p \times g_{\rm max}(1+\delta_{ij})$, with $p=0$ (grey dots), $0.125$   (grey dash dots), $0.15625$ (grey dashes), $0.25$ (grey solid), $1$ (dark grey), $5$ (black), and $10$ (purple).}
\label{fig9}
\end{figure*}

Intracellular recordings from the four identified interneurons of the swimming CPG of the {\it Melibe} have indicated that the phase-lags, 
($\Delta \phi_{12},\Delta \phi_{13},\Delta \phi_{14}$), between the burst initiation in the voltage traces are maintained stably at these values: ($1/2,\,3/4,\,1/4$). While it is evident that the both HCOs always remain bursting in anti-phase, it is less clear what mechanisms, involving reciprocal inhibition and/or excitation, polarity of wiring etc., are used by 4-neuron networks to preserve the stability of $3/4$ phase-lag between the HCOs. Assuming that the building blocks of such networks remain HCOs formed by anti-phase bursting interneurons,  our next step in exploration of such networks, is the examination of functions of contralateral inhibition (circuit in Fig.~\ref{sfig17a}) and ipsilateral excitation (the circuit shown in Fig.~\ref{sfig18a}), and whether they can break or enforce the robustness of activity patterns. 
In what follows, we explore the dependence of the phase-lag between the HCOs, i.e.  $\Delta \phi_{13}$, or equivalently, $\Delta \phi_{14}$, on the coupling strengths in heterogeneous networks.  Panels of Fig.~\ref{fig9} shows how $\Delta \phi_{13}$ varies as unidirectional feed-forward inhibition from HCO2 onto HCO1, and unidirectional backward excitation from HCO1 onto HCO2  are increased. Since both neural configurations are sub-circuits of the {\it Melibe} CPG, let us consider them independently first.

A phase-locked network state turns out to be quantitatively a function of unidirectional inhibition and excitation coupling. When the inhibitory  strength, $g_{32}^{\rm inh}=g_{41}^{\rm inh}$,  of the contralateral synapses is increased, the phase-locked state, initially emerging at low small values of $\Delta \phi_{13}$,  quickly moves to a high value around $0.9$, see Fig.~\ref{sfig17}. This means that on average,  interneuron~3, first following interneuron~1, becomes delayed by nearly a bursting period as the inhibition between the HCOs substantially increased.    

In the case of ipsilateral excitation, increasing  $g_{13}^{\rm exc}=g_{24}^{\rm exc}$  works the other way around for synchronization of the interneurons: 1 with 3, and 2 with 4. The evolution of the steady states of the phase-lag, $\Delta \phi_{13}$ is presented in Fig.~\ref{sfig18}. It shows that above a threshold $g_{\rm exc}=0.5\,g^{\rm max}$, the phase-lag, $\Delta \phi_{13}$, shifts down to a steady state, i.e. the driven interneuron 3 (4) follows interneuron 1 (2) after a short delay of $1/10$-th of the period of the network, and so does HCO2 after HCO1, as a whole.    

Given that phase-lags are defined on modulo 1, one can say that in-phase synchrony between the HCOs is due to repulsion in the case of contralateral inhibition, and due to attraction in the excitatory case. A simple calculation (given in appendix) demonstrates that for the network to achieve a robust phase-locked state, the driven HCO has to adjust its period, i.e. either catch up or slow down, in order to match up with that of the driving HCO in unidirectional cases.  
The effect of increasing synaptic strengths saturates in both cases, after some thresholds are reached. Making the coupling strength five times stronger than the nominal value of the maximal synaptic conductance, has little effect (purple lines in Figs.\ref{sfig17} and \ref{sfig18}) on the steady state value of $\Delta \phi_{13}$.   Comparison of two types of coupling (network configurations) suggests that the contralateral inhibition produces a phase-locked state that appears to be the closest to the experimentally observed pattern. 

Because there are other, excitatory and electrical connections between the interneurons in the CPG circuitries, in the following sections we will address and identify their roles for predominance and robustness of specific bursting states in 4-cell CPG networks. It is shown in \cite{NSLGK012} that while the swim CPG of {\it Melibe} with strong contralateral inhibitory synapses produces patterns with high $\Delta \phi_{13}$ values, the swim CPG of a another sea slug {\it Dendronotus}, possessing ipsilateral excitatory connections, produces bursting patterns with low $\Delta \phi_{13}$ values, which agrees with our findings.

\subsection{Modulatory effect of electrical coupling}

\begin{figure*}[!htbp]
\centering
\subfigure [][]  {\resizebox*{.47\columnwidth}{!}{\includegraphics{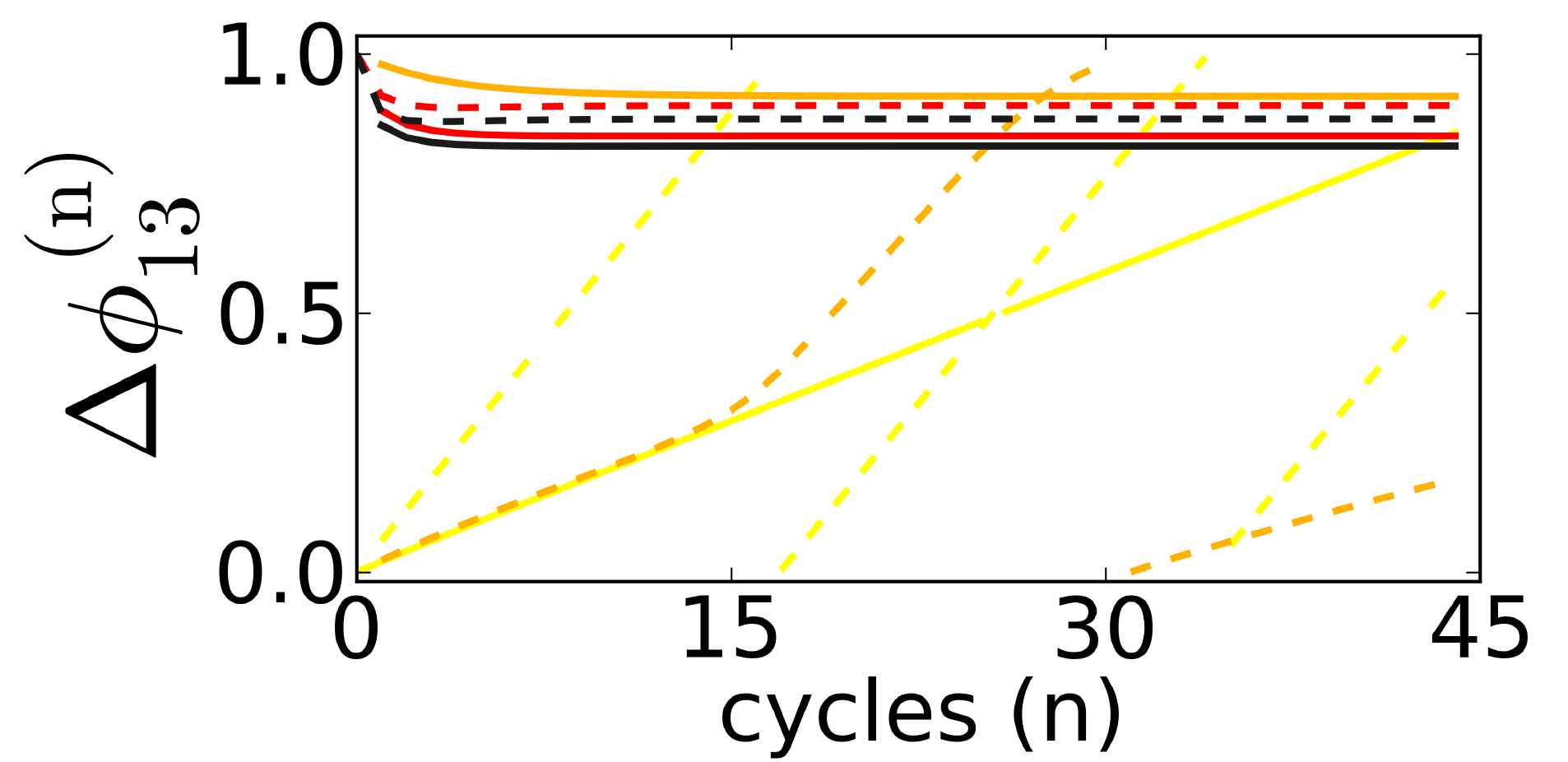}}\label{sfig10a}}
\subfigure [][]  {\resizebox*{.47\columnwidth}{!}{\includegraphics{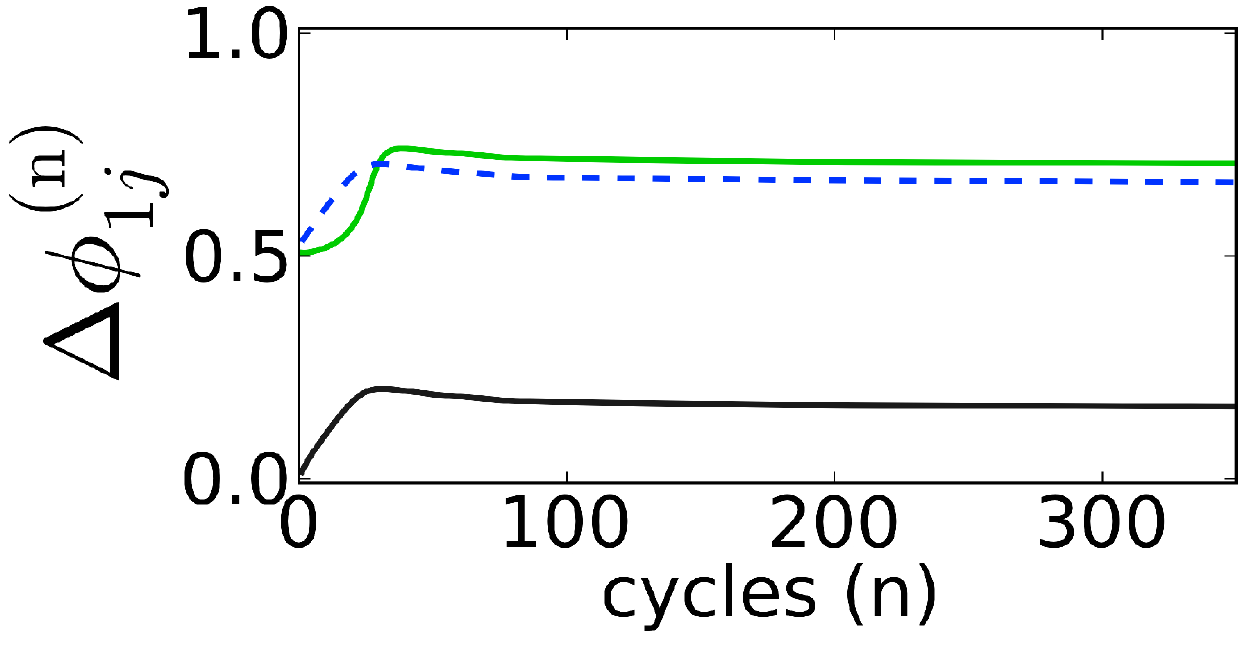}}\label{sfig10b}}
\caption{(a) Transformations of the steady states of of $\Delta \phi_{13}$ phase-lag in the network in Fig.~\ref{sfig17}  the contralaterally inhibitory coupling is increased: 
$g_{32}^{\rm inh}=g_{41}^{\rm inh}=d \times g_{\rm max}\,(1+\delta_{ij})$, with $d=0$ (yellow curve), 
$1$ (orange), $5$ (red) , and  $10$ (black); $g^{\rm elec}=0.5\,g_{\rm max}$ and $g_{\rm max}=2.5 \times 10^{-3}$. 
(b) Progression of phase-lags, $\Delta \phi_{12}$ (green), $\Delta \phi_{13}$ (black) and $\Delta \phi_{13}$ (dashed blue) to the steady state,  plotted against the burst cycle number for 
$g_{32}^{\rm inh}=g_{41}^{\rm inh}=1.25 \times g_{\rm max}\,(1+\delta_{ij})$.} \label{fig10}
\end{figure*}

Electrical coupling, or gap junctions, provide bidirectionally a  continuous interaction between interneurons thus affecting synchronization properties of oscillatory neural networks \cite{SZVC99}. 
 Its magnitude, proportional 
to the difference between the current values of the membrane potentials, promote in-phase synchronization, in most cases.  We introduce electrical coupling (represented by a resistor in the circuitry in Fig.~\ref{sfig3a}) between interneurons 1 and 2 of HCO1 in the form  $g^{\rm elec}\, (V_1-V_2)$ and visa versa, 
 in addition to the contralateral  inhibition from HCO2 to HCO1 in the heterogeneous network presented in the previous section.
While we refer, in general, to an HCO (HCO1 here) as a pair of interneurons bursting in alternation,  a proper gap junctions can overcome 
the inhibition-caused anti-phase dynamics and synchronize the interneurons to burst as a whole.  Prior to that, for values of $g^{\rm elec}$ below a synchronization threshold,  the period of HCO1 will
gradually vary with an increase of the electrical coupling strength, leading to changes in phase-locked states and transformations of bursting patterns of the CPG, as depicted in Fig.~\ref{sfig10a}.

Figure \ref{sfig10b} shows how the phase-lags, $\Delta \phi_{1j}$, of the CPG change with an increase of the electrical synapse through 
$g^{\rm elec}=0.5g_{\rm max}$ in HCO1. The gap between interneurons 1 and 2 widens to $0.75$ (toward synchrony at $1$), as they keep receiving the contralateral  inhibition from interneurons 4 and 3, bursting in anti-phase:  $\Delta \phi_{34}=\Delta \phi_{14}-\Delta \phi_{13}=1/2$, as  the diagram suggests.

\section{Range of heterogeneity of the CPG} \label{range_heterogeneous}

\begin{figure*}[!htbp]
  \centering
  \subfigure [][] {\resizebox*{.2\columnwidth}{!}{\includegraphics{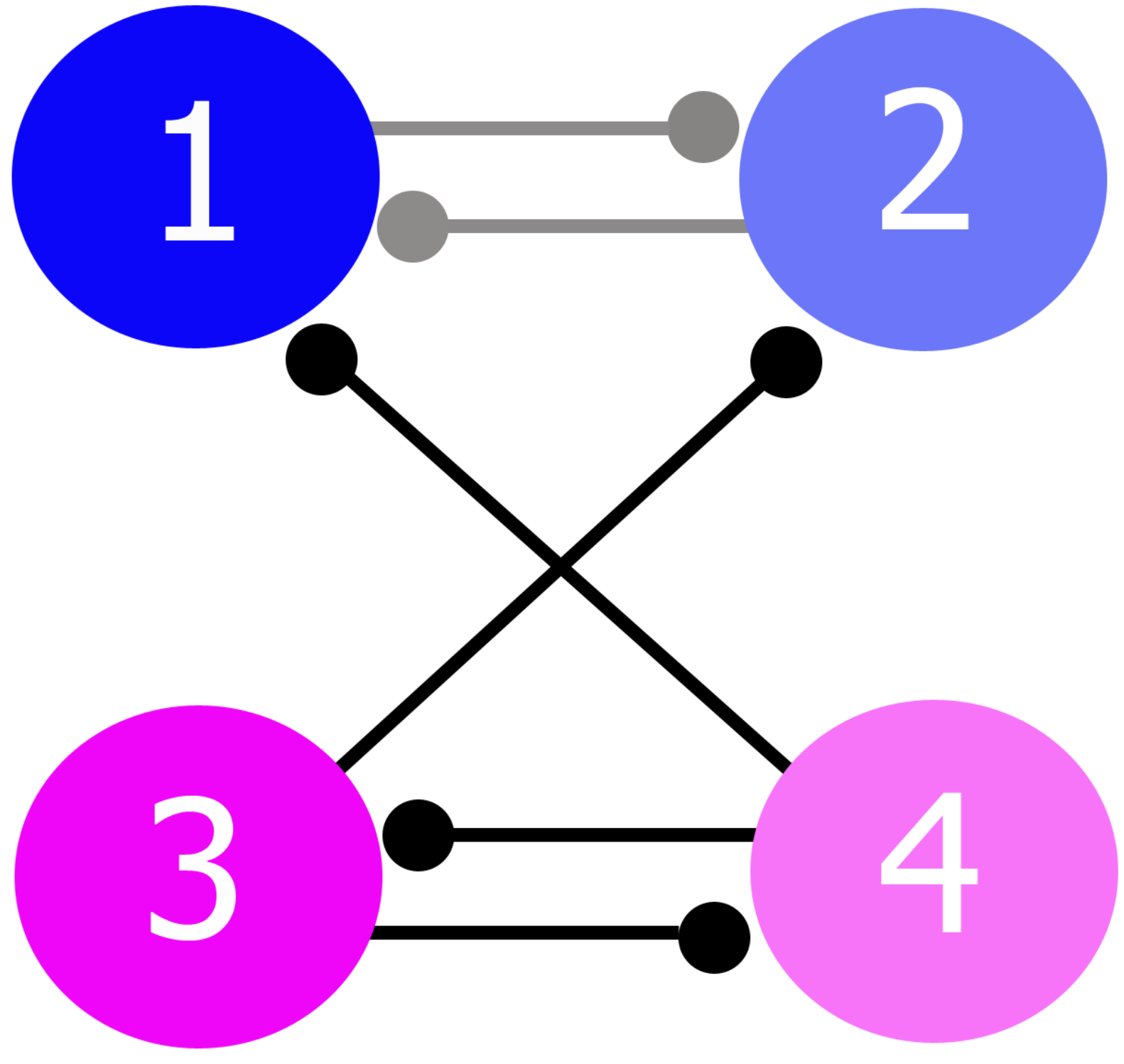}} \label{sfig19a}}~~~~~~~
  \subfigure [][] {\resizebox*{.2\columnwidth}{!}{\includegraphics{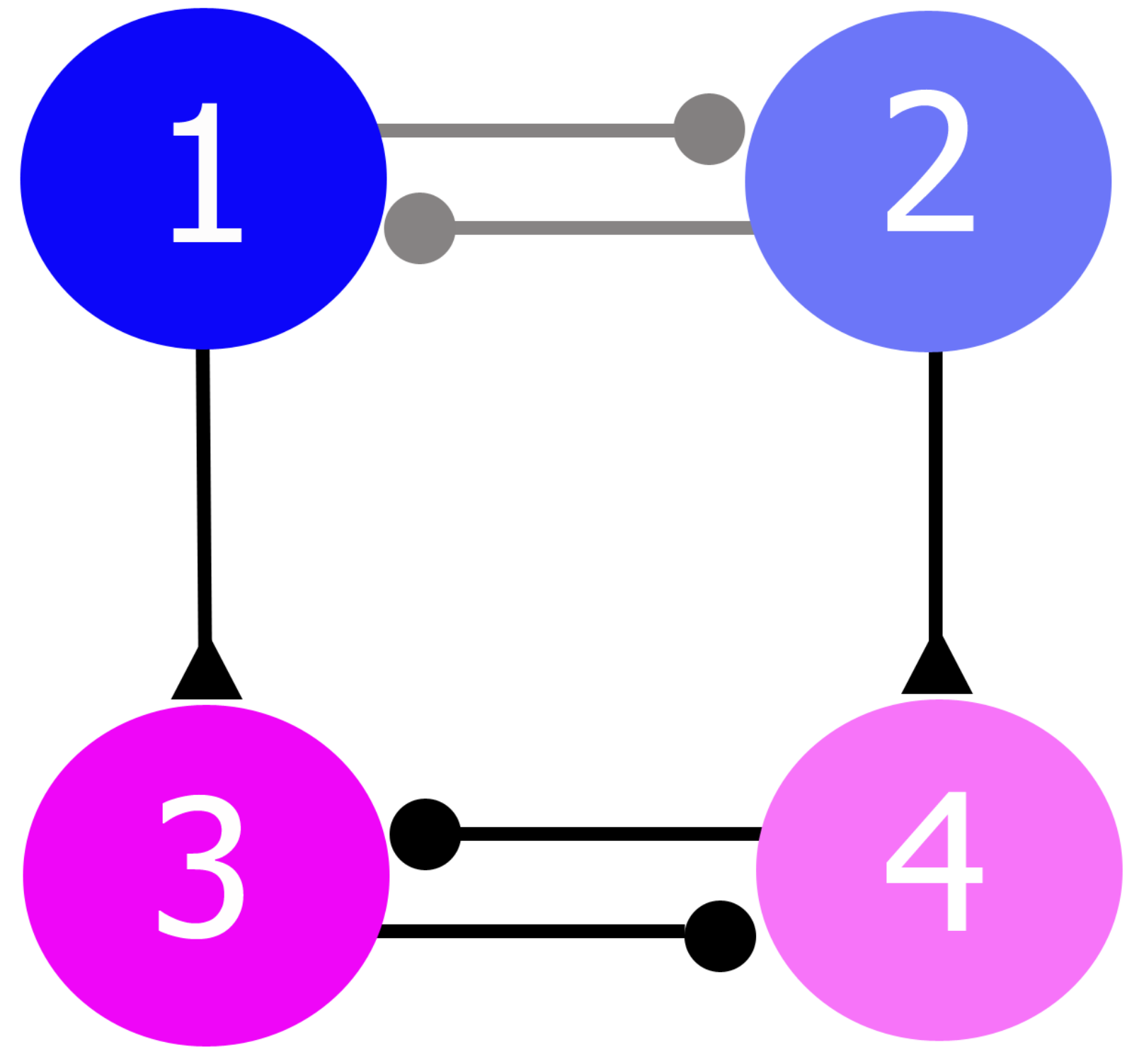}}\label{sfig20a}}~~~~~~~
  \subfigure [][] {\resizebox*{.2\columnwidth}{!}{\includegraphics{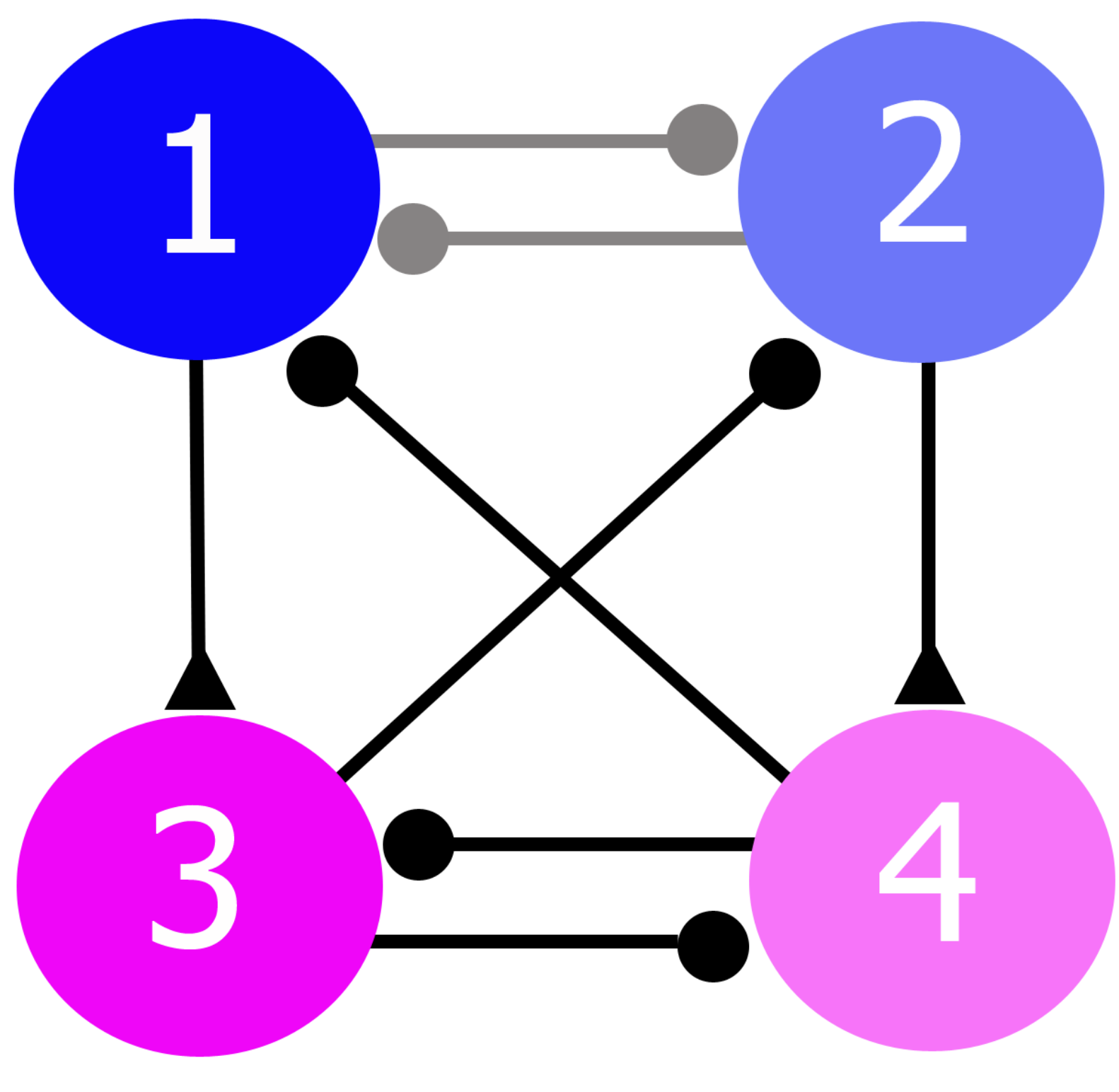}}\label{sfig21a}}~~~~~~~
  \subfigure [][] {\resizebox*{.2\columnwidth}{!}{\includegraphics{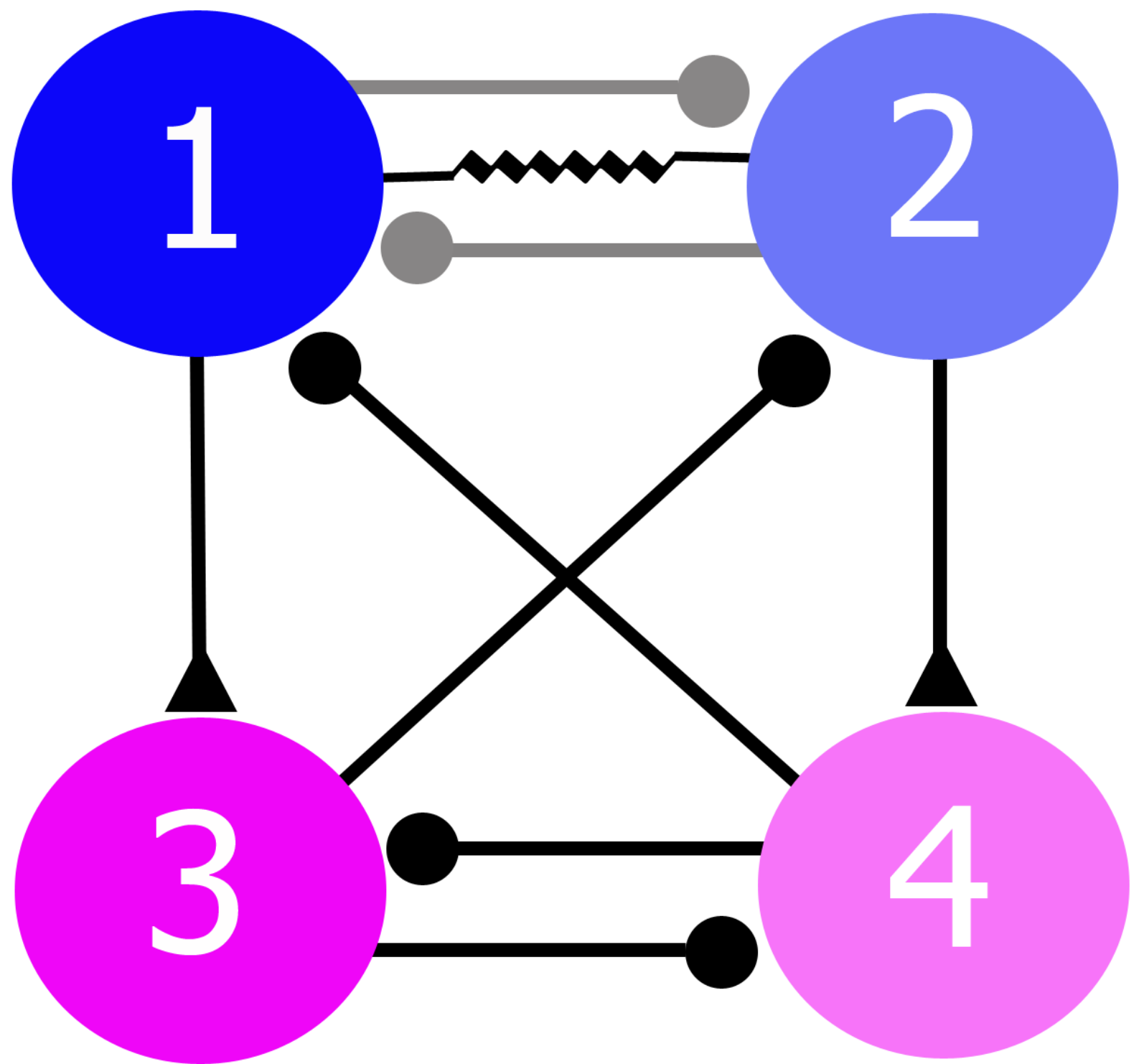}}\label{sfig22a}}
  \caption{Four circuitries of coupled HCO networks being tested for the heterogeneity range: (a) contralaterally inhibitory (I) synapses (denoted by round-headed arrows); (b) ipsilaterally excitatory (E) synapses denoted by triangle-headed arrows; (c) mixed: with contralaterally inhibitory and ipsilaterally excitatory synapses (IE); (d) the complete CPG with a gap junction or electrical synapse (IEG) (denoted by a resistor symbol)  between interneurons 1 and 2.}
  \label{fig11}
\end{figure*}

We have pointed out earlier that weakening reciprocal inhibition between interneurons of one HCO can be equivalent to 
strengthening reciprocal inhibition in the counterpart. In this section, we examine the 
range of heterogeneity of the 4-cell network in terms of the misbalance among the synaptic coupling strengths. We will vary the reciprocal
inhibitory coupling in HCO1 only, while having those in HCO2 intact along with the contralateral, ipsilateral ($g_{\rm max}=2.5 \times 10^{-3}$) and electrical ( $g^{\rm elec}=0.25\,g_{\rm max}$) connections.  The following four network configurations, schematically drawn in Fig.~\ref{fig11}, are explored in this section. In addition to unidirectional cases, we combine them,  in the mixed CPG (Fig.~\ref{sfig21a}), in which the gap junction will next bridge the interneurons of HCO1  (Fig.~\ref{sfig22a})

\begin{figure*}[!htbp]
  \centering
  \subfigure [][] {\resizebox*{.47\columnwidth}{!}{\includegraphics{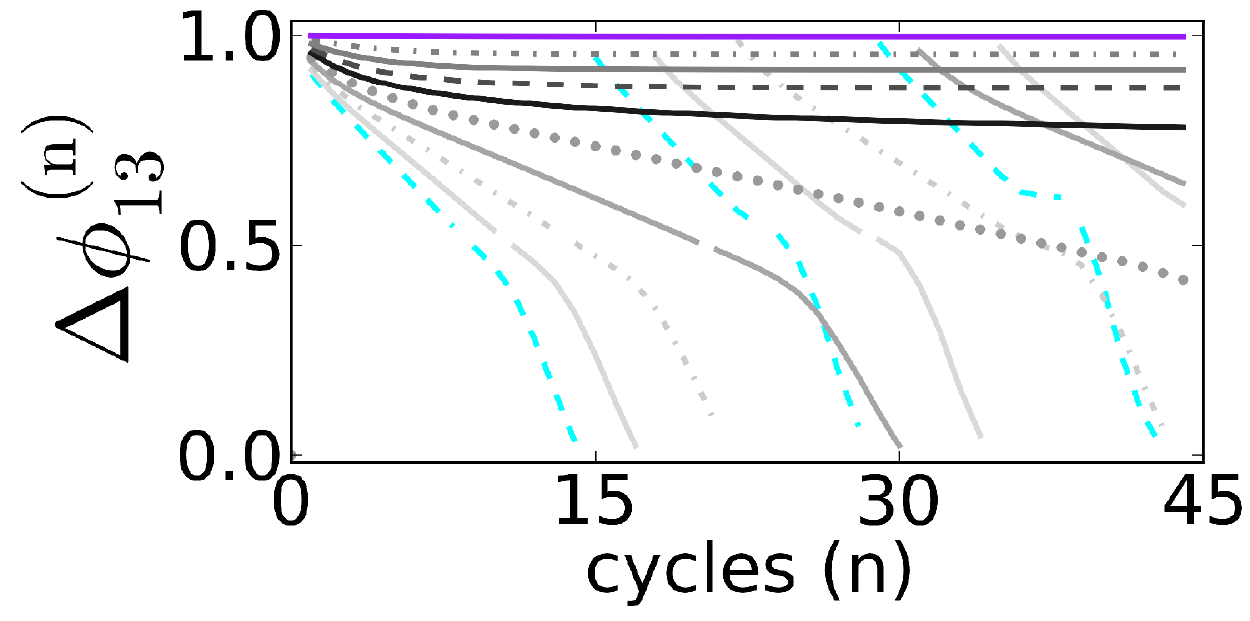}} \label{sfig19}}
  \subfigure [][] {\resizebox*{.47\columnwidth}{!}{\includegraphics{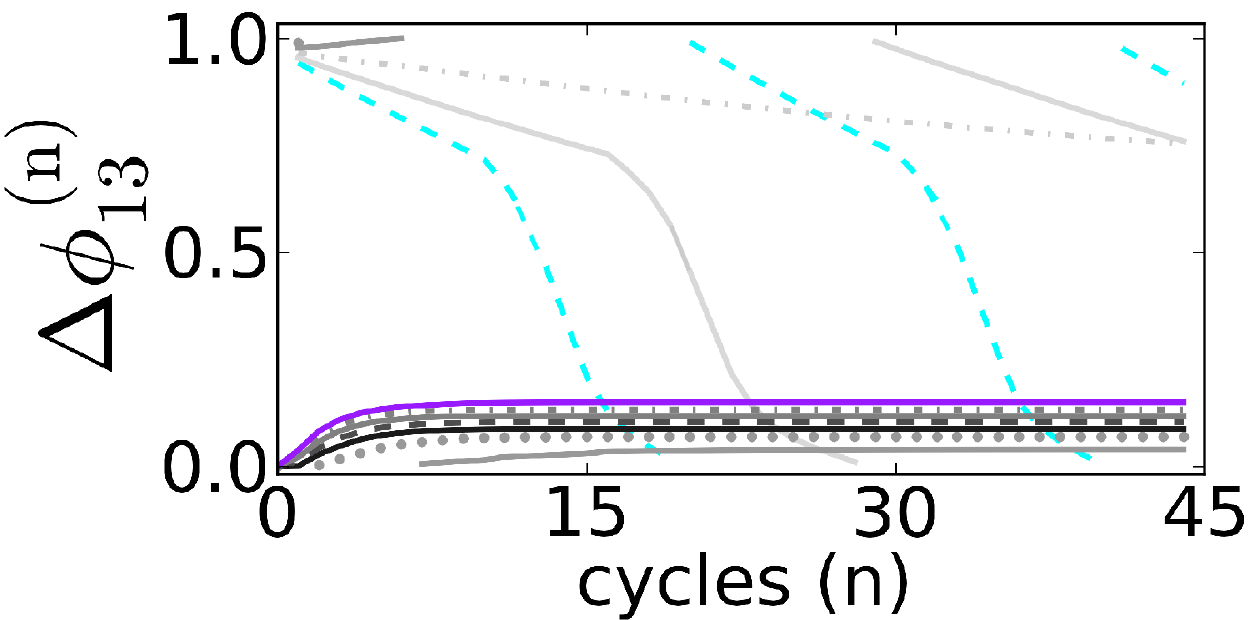}} \label{sfig20}}\\
  \subfigure [][] {\resizebox*{.47\columnwidth}{!}{\includegraphics{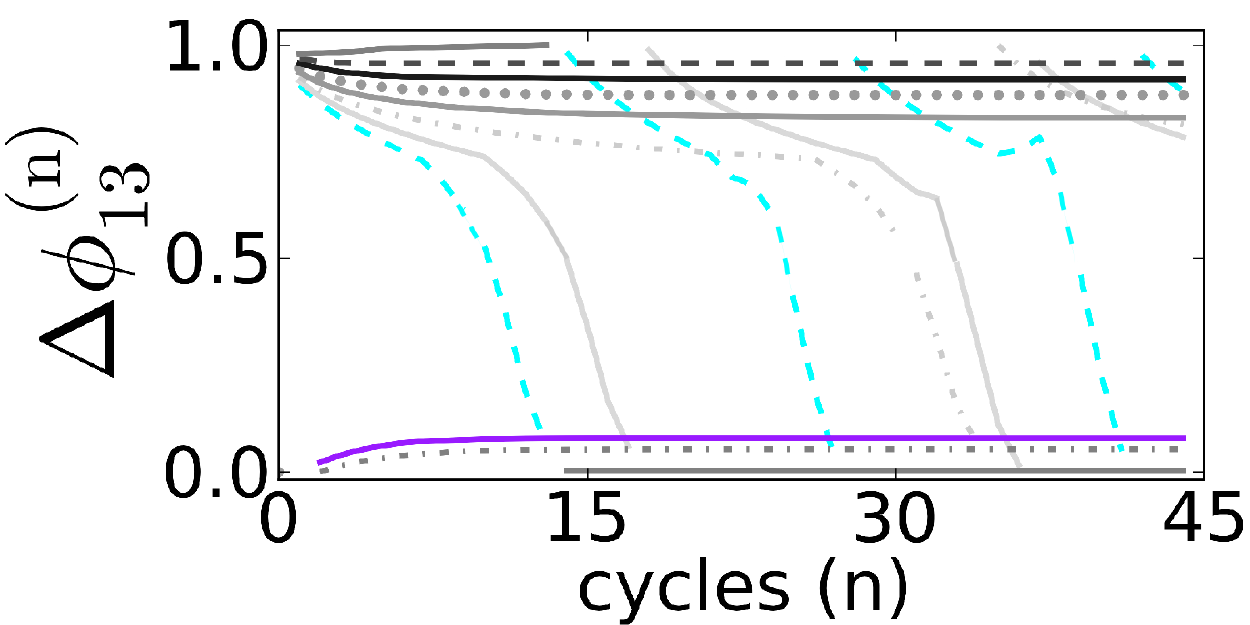}}\label{sfig21}}
  \subfigure [][] {\resizebox*{.47\columnwidth}{!}{\includegraphics{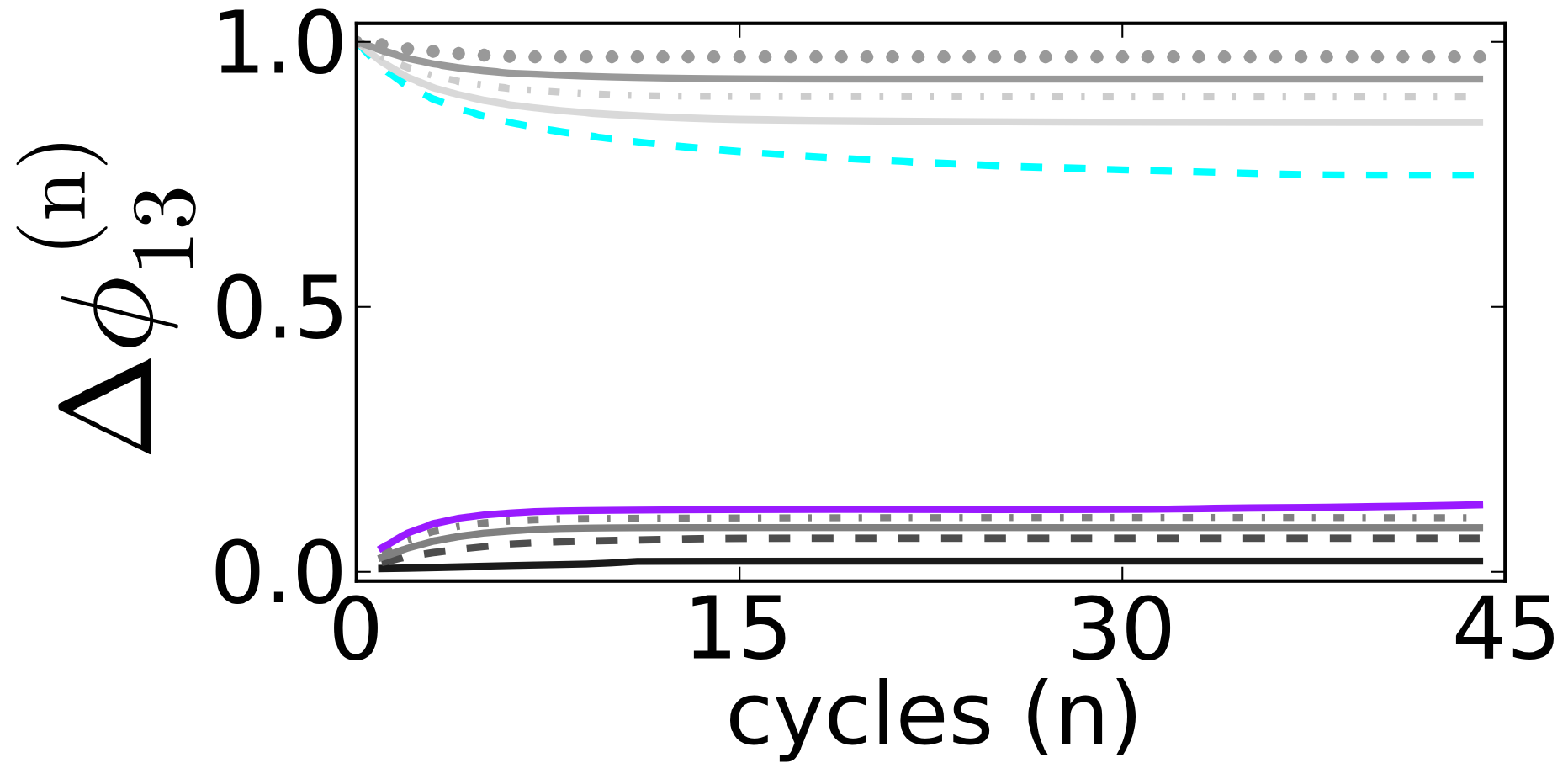}}\label{sfig22}}
  \caption{ Four panels (a-c) showing 45-cycles long transients of the $\Delta \phi_{13}$ phase-lags for the four corresponding CPG circuits: I, E, IE and IEG, respectively (Fig.~\ref{fig11}(a-c)) as the HCO1 inhibitory coupling, $g_{12}^{\rm inh}=g_{21}^{\rm inh}=b \times g_{\rm max}$, is increased: $b=0$ (purple curve); $b=0.25$ (grey dash dots); $0.5$ (grey solid); $0.75$ (dark grey dash), $1$ (black); $1.25$ (grey dots); $1.5$ (light grey long dash/solid); $1.75$ (light grey dash-dots); $2$ (light grey long dash/solid); and $2.25$ (cyan curve). Here, 
%$g_{34}^{\rm inh}=g_{43}^{\rm inh}=g_{32}^{\rm inh}=g_{41}^{\rm inh}=g_{\rm max}(1+\delta_{ij})$, $g_{13}^{\rm exc}=g_{24}^{\rm exc}=g_{\rm max}$,
$g_{12}^{\rm elec}=g_{21}^{\rm elec}=0.25g_{\rm max}$ and $g_{\rm max}=2.5 \times 10^{-3}\,(1+\delta_{ij})$.}
  \label{fig12}
\end{figure*}

A robust phase-locked state of a bursting pattern in a network must persist for a certain range in the high-dimensional parameter space of the coupling weights.  Given a large number parameters in a generic 4-cell network with various connections, we need to come up with reduction assumptions  to single out one effective control parameter, while other less principle parameters are kept fixed. As such an effective control parameter, we employ  the ratio of the inhibitory  strengths in individual HCOs. We will start with the case of nearly uncoupled interneurons 
in HCO1 at small $g^{\rm inh}_{12}=g^{\rm inh}_{21}$, next the reciprocal inhibition is increased to the nominal value, $g_{\rm max}$, and then 
made 2.5 stronger than $g^{\rm inh}_{34}=g^{\rm inh}_{43}$.  The evolution of the representing phase-lag, $\Delta \phi_{13}$, is presented in 
Fig.~\ref{fig12} for the four network configurations.  In the diagram, the purple, black and cyan curves correspond to a representative trajectory converging to an unique attractor as we increase inhibition 
in HCO1. We note that in all cases in question, both HCOs remain anti-phase bursters, i.e. $\Delta \phi_{12}=\Delta \phi_{34}=1/2$, in the network, so variations in coupling can only shift the phase-lag between them.  

We can conclude from the examination of the convergence tendencies of $\Delta \phi_{13}$ that a phase-locked state exists in all cases. The widest range of heterogeneity is observed for the feedback configuration in Fig.~\ref{sfig22} with the regulatory gap junction.  Furthermore, all but one configuration can produce a stable network state with the desired phase-lag 
$\Delta \phi_{13}=3/4$. 

\begin{figure*}[!htbp]
  \centering
  {\resizebox*{.5\columnwidth}{!}{\includegraphics{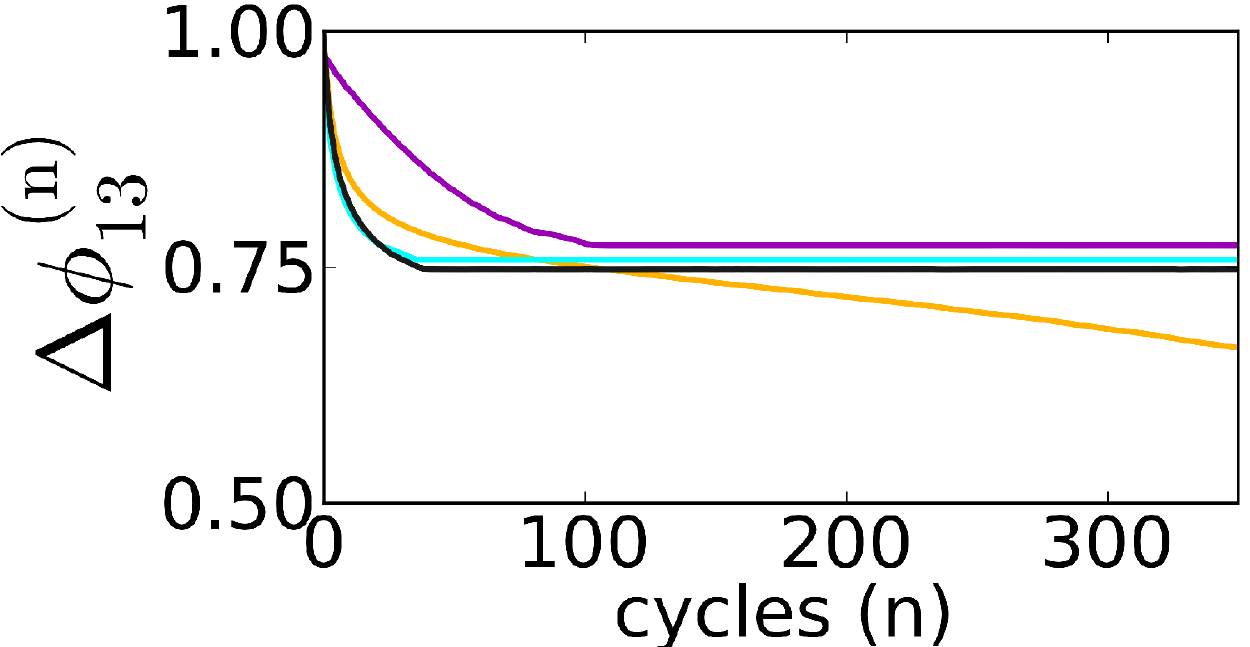}}}
  \caption{Convergence of the $\Delta \phi_{13}$ phase-lags to the steady state at $3/4$ after 350 burst cycles for the four corresponding CPG circuits in Fig.~\ref{fig11}: yellow (I) , purple (E), cyan (IE) and black (IEG) lines, respectively. 
  Cyan curve corresponds to the following conductances: $g^{\rm inn}_{34}=g^{\rm inh}_{43} = 2.25g_{\rm max}$ and $g^{\rm elec}_{12}=g^{\rm elec}_{21} = 0.25g_{\rm max}$, with the rest at the nominal value. } \label{fig13}
\end{figure*}

For the contralaterally inhibitory configuration (I), sketched in Fig.~\ref{sfig19a}, to have $\Delta \phi_{13}=3/4$, the reciprocal inhibitions within HCO1 must be weaker than in HCO2. Having it equal or stronger, leads to the loss of phase-locked state (transient black and cyan lines in Fig.~\ref{sfig19}).  Fig.~\ref{fig13} presents the outcome of simulations of the network dynamics for longer traces over 350 burst cycles. In this diagram, the only $\Delta \phi_{13}$ transient (yellow line)   for the inhibitory homogeneous CPG (equal coupling) shows no stabilization.   

For the configuration (E) with ipsilaterally excitatory synapses between the HCOs (Fig.~\ref{sfig20a}), short-term simulations (45 burst cycles) shows no phase-locking even for the large values of $g_{12}^{\rm inh}=g_{21}^{\rm inh}$, however continuation simulations over 200 bursting cycles indicate a slow convergence (purple curve in Fig.~\ref{fig13}) to the phase-lag  $\Delta \phi_{13}=3/4$. 

We can conclude that for the CPG model to maintain robustly and flexibly the desired phase-locking at $(\Delta \phi_{12},\,\Delta \phi_{13},\,\Delta_{14})=(1/2,\,3/4,\,1/4)$ like that in the {\it Melibe} swim CPG, all connections, contralateral inhibitory, ipsilateral excitatory,  and 
electrical, are necessary. These connections provide the  feedback loop that widens the range of heterogeneity, within which the CPG network possesses the phase-locked state corresponding to the swim bursting pattern. Another related observation suggests that HCO1 should generate reciprocally inhibition stronger than HCO2 to preserve the locking balance. Finally, addition of relatively strong electrical coupling modulates the 
phase-locked state (black curve in Fig.~\ref{sfig22}), that allows the heterogeneous CPG with quite distinct HCOs to maintain the stable 
phase-lags at $(1/2,\,3/4,\,1/4)$.

\subsection{Reduced maps for the phase-lags between HCOs.}

Oscillatory network states and their transformations can be effectively identified and studied through the use of 
Poincar\'e return maps. In this section, we use the maps to examine a particular bursting pattern of the inhibitory 4-cell 
network (in Fig.~\ref{sfig19a})  that corresponds to the in-phase bursting interneurons of the driving HCO2: $\Delta \phi_{34}=0$.  In the 2D  
$(\Delta \phi_{13}, \Delta \phi_{14})$ phase-lag  projection, this pattern is associated with the solutions belonging to the main 
(red) diagonal in Fig.~\ref{sfig15}. The diagonal is indeed an invariant plane inside
the unit cube on which the return for all three phase-lags is defined. 
In restriction to this plane, the 4-cell network is reduced  to the 3-cell one, in which 
 the anti-phase bursting interneurons of HCO1 receive double inhibition during the active phase of the in-phase bursting 
HCO2. For the reduced network, the return map becomes a two-dimensional map defined on 
the phase-lags, $\Delta \phi_{12}$ and $ \Delta \phi_{13} \equiv \Delta \phi_{14}$.  

\begin{figure*}[!htbp]
  \centering
  \subfigure [][] {\resizebox*{.45\columnwidth}{!}{\includegraphics{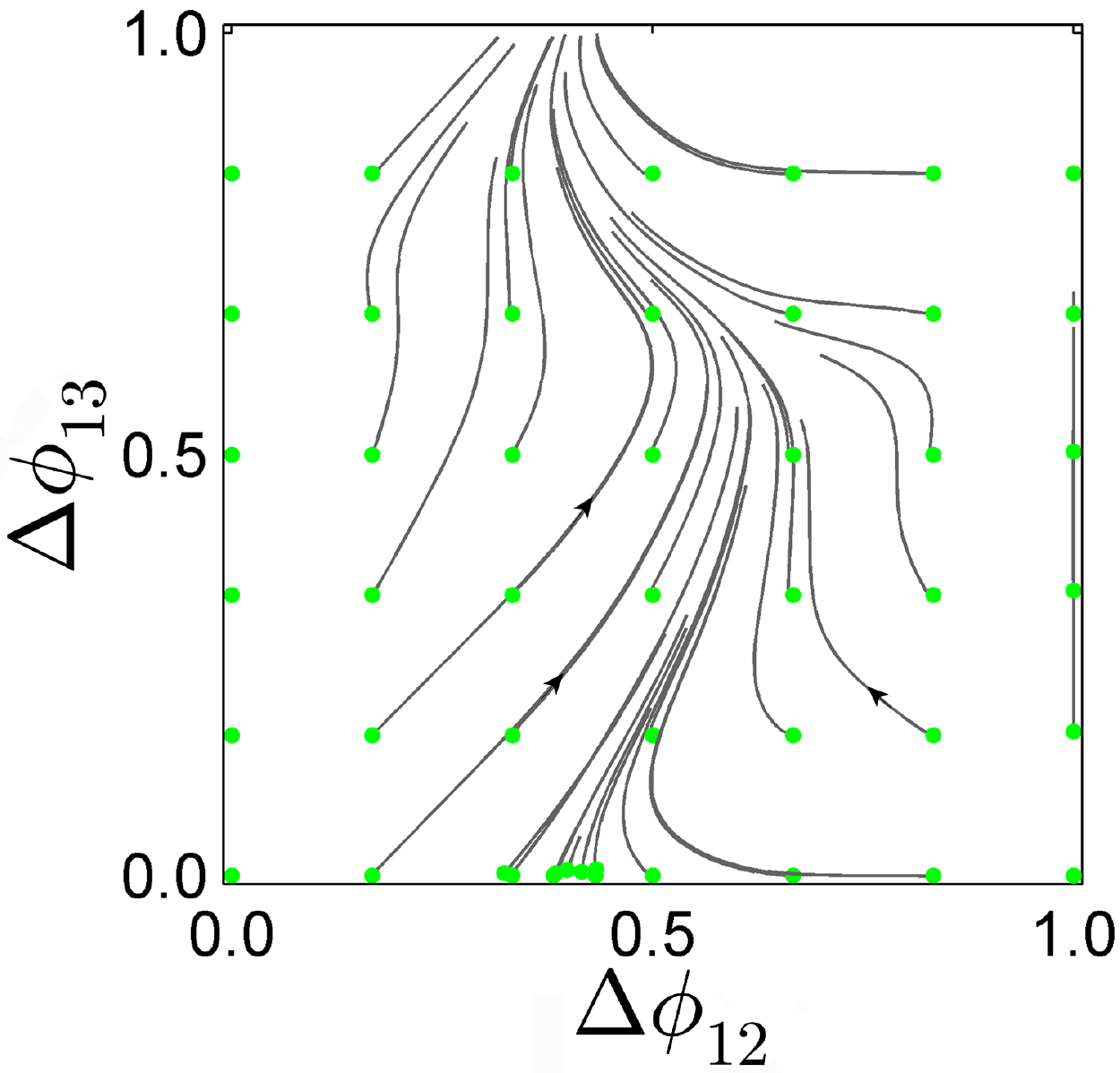}} \label{sfig23}}
  \subfigure [][] {\resizebox*{.47\columnwidth}{!}{\includegraphics{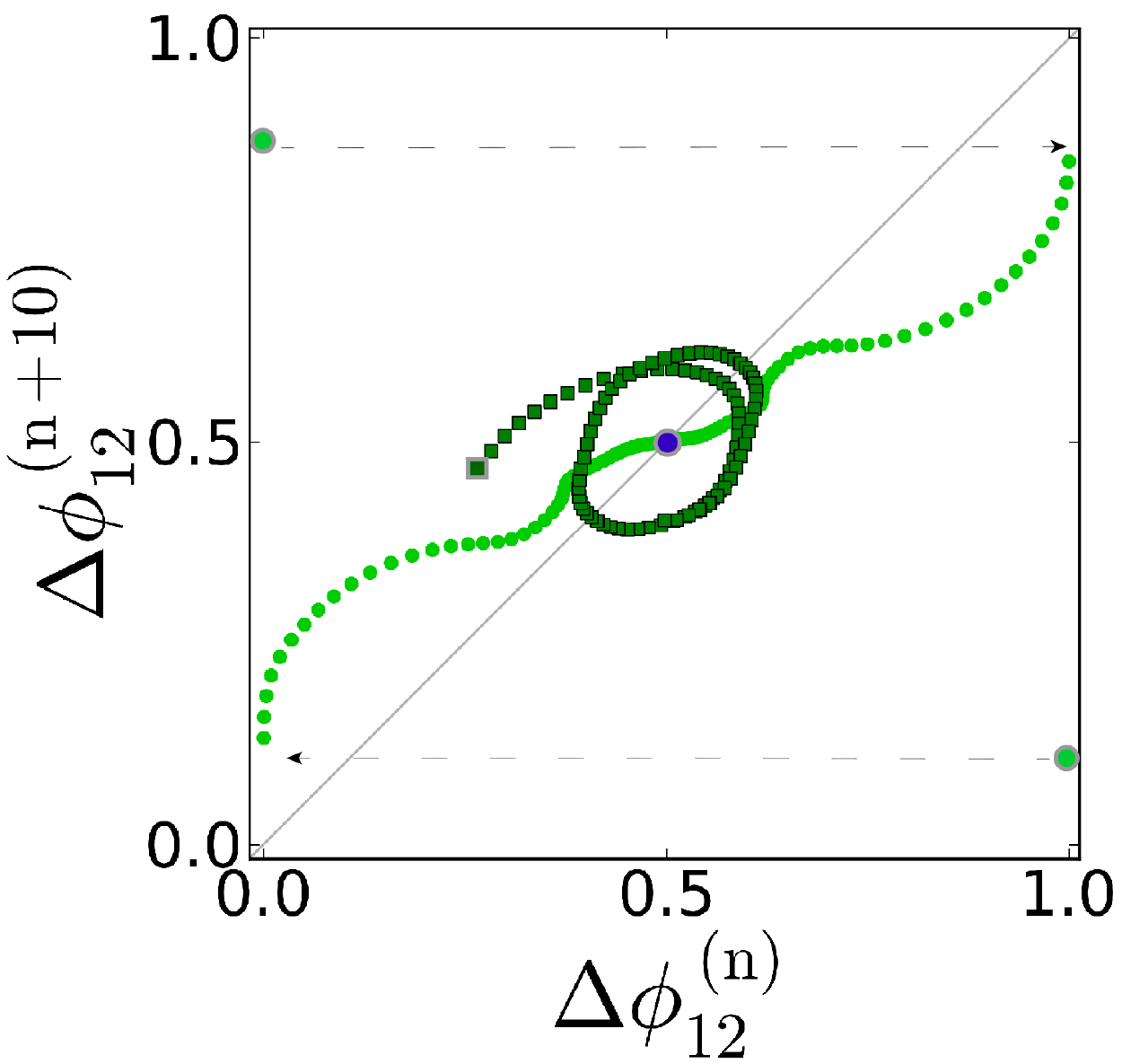}} \label{sfig24}}\\
  \caption{2D ($\Delta \phi_{12}, \,\Delta \phi_{13}$) phase-lag map  for the inhibitory 4-cell CPG with the in-phase bursting HCO2 
forcing periodically  the $\Delta \phi_{12}$ phase-lag for HCO1.  
(a) Trajectories, subjected to the constrain  $\Delta \phi_{13} = \Delta \phi_{14}$, of the map
for $g_{12}=g_{21}=g_{32}=g_{41}=5 \times 10^{-4}$ and $g_{12}=g_{21}=10^{-3}$, converging to an invariant curve wrapping around the unit square 
(2D torus).  Green dots represent initial phase-lags. (b) 1D return map: 
$\Delta \phi_{12}^{n} \to \Delta \phi_{12}^{n+10}$:  a stable invariant circle (dark squares) is sampled from the solutions in (a);   
 light dots represent typical, unconstrained trajectories of the map converging to the fixed point at $\Delta \phi_{12} = 0.5$ on 
 the $45^\circ$-line for $g = 10^{-3}$.}  \label{fig14}
\end{figure*}

The dynamics of such a map can be assessed by following  forward transients (grey) shown in Fig.~\ref{sfig23}, 
 whose initial conditions (green dots) are subjected to the synchronization condition $\Delta \phi_{13} = \Delta \phi_{14}$.  
The return map reveals a stable invariant curve wrapping around the unit square (2D torus). In the absence of fixed points,  the torus must 
contain a matching unstable invariant curve too.
Because it is unstable and repels forward iterates of the map, we may hypothesize that it wiggles around the unstable in-phase state, $\Delta \phi_{12}=0$, of HCO1. This state is unstable because of breaking perturbations due to periodic forcing originated from HCO2.  

In essence,  the in-phase bursting HCO2 periodically drives or modulates the phase-lag, $\Delta \phi_{12}$,  
causing the onset of  oscillations, or phase jittering, around  $1/2$ corresponding to the anti-phase bursting HCO1.  
This observation lets us define  a further reduced 1D map for the discrete evolution of the phase-lag between the reference and any other 
interneurons of the CPG: $\Delta \phi_{1j}^{(n)} \to \Delta \phi_{1j}^{(n+k)}$, where $k$ is the degree of the map. The map 
$\Delta \phi_{12}^{(n)} \to \Delta \phi_{12}^{(n+10)}$ for HCO1 is shown in Fig.~\ref{sfig24};  $k=10$ is chosen because of the slow convergence of transients to an attractor in the weakly coupled case.  
In this figure, the modulation oscillations of $\Delta \phi_{12}$ are represented by
the closed invariant circle (dark green dots). Such an invariant circle can be made of finite or infinite number of points depending on whether the ratio of the burst periods of the HCOs is rational or not.  This circle in the 1D map corresponds to the stable invariant curve 
wrapping around the 2D torus in Fig.~\ref{sfig23}. In the case where the interneurons of  HCO2 bursts in anti-phase,  $\Delta \phi_{12}$ phase-lag transients converge monotonically to the fixed point at the intersection of the map graph (light green) curve with the $45^\circ$-line. The fixed point corresponds to the anti-phase bursting interneurons 1 and 2 of HCO1. An unstable fixed point of the 1D map at the origin (at 1) 
corresponds to the repelling invariant curve in the 2D map in Fig.~\ref{sfig23}.

\begin{figure*}[!htbp]
  \centering
  \subfigure [][] {\resizebox*{.4\columnwidth}{!}{\includegraphics{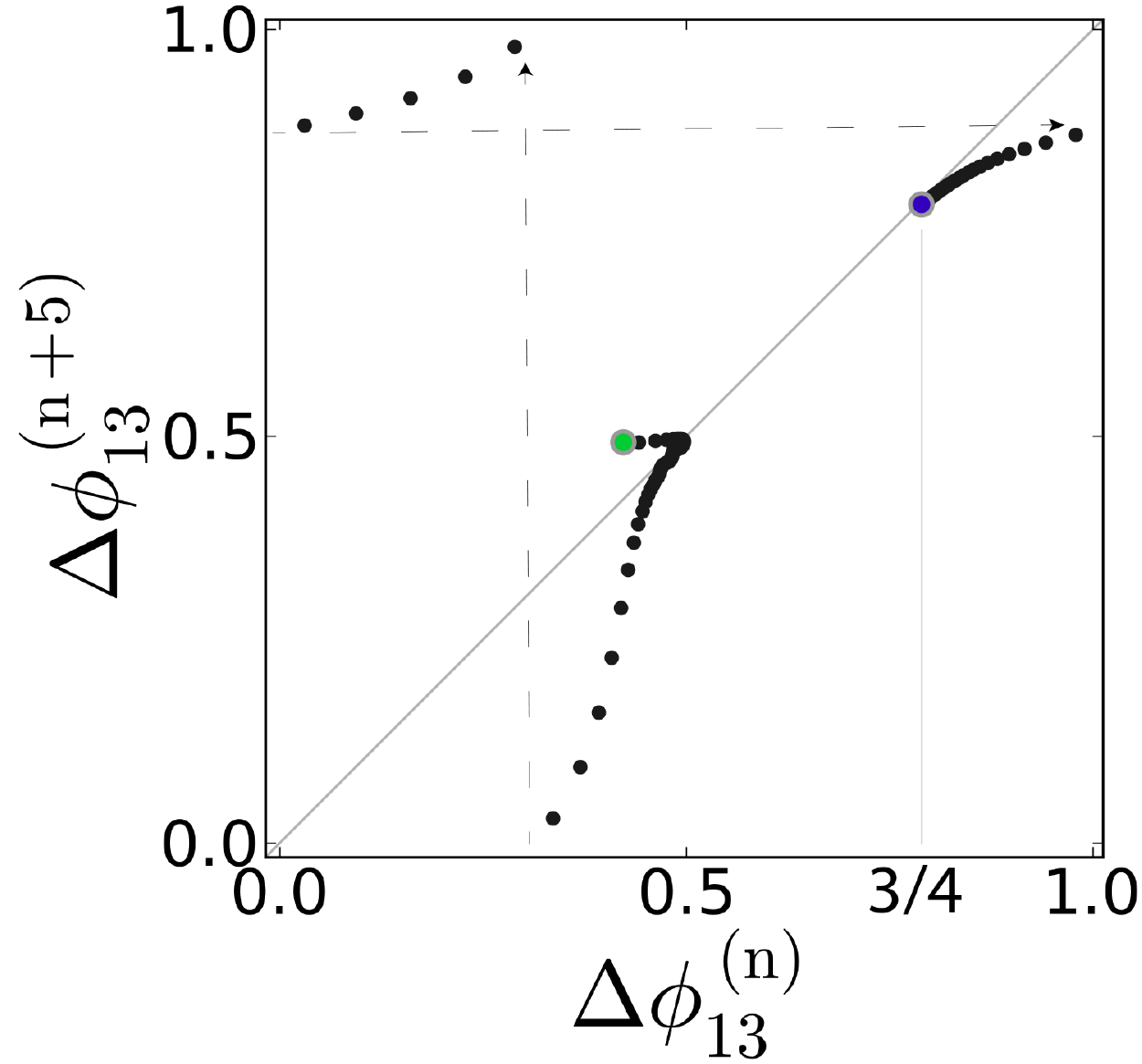}} \label{sfig25}}
  \subfigure [][] {\resizebox*{.4\columnwidth}{!}{\includegraphics{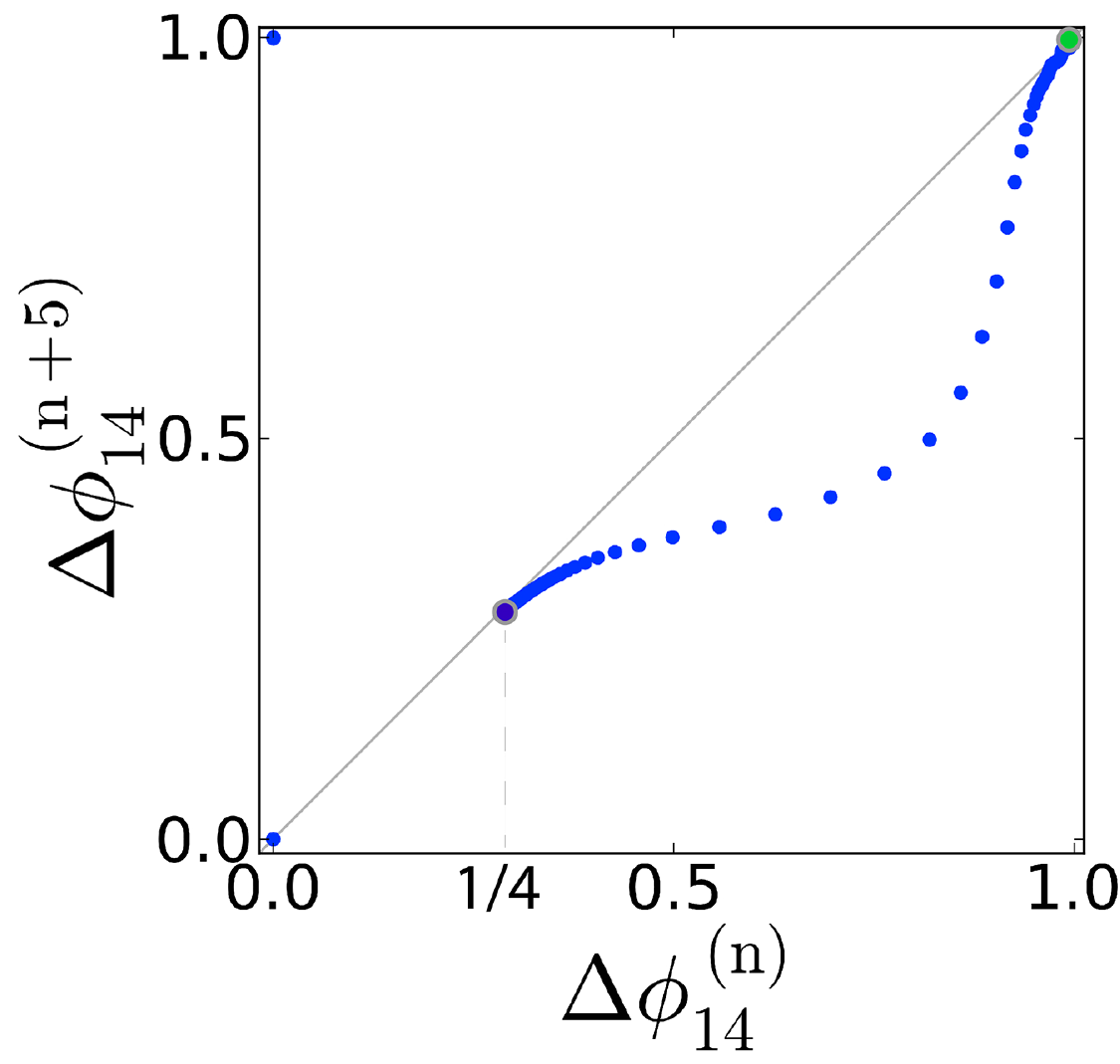}} \label{sfig26}}\\
  \caption{1D  maps for the phase lags, $\Delta \phi_{13}$ (a), and $\Delta \phi_{14}$ (b), showing the complex ways of the convergence of the 
iterates  to the fixed point with the coordinates  $3/4$ and $1/4$, respectively, corresponding to the single stable pattern of the 
inhibitory CPG. On the route toward the point(s), the transients slow down near a saddle represented 
 by an unstable fixed point with the coordinates $1/2$ and $1$, respectively, in the panels. }
 \label{fig15}
\end{figure*}

Due to the periodic nature of the patterns in this study, we employ return maps to investigate the ways transients converge to attracting states 
such as fixed points and invariant circles.  Each phase-lag trajectories form a distinct discrete path on the 3D torus (unless it is periodic on 
an invariant curve). While  reducing to 1D return maps is practical in many instances for detecting stable fixed points for phase-locked states of the network, interpretation of solutions of reduced maps can be ill-suited for proper description of high-dimensional dynamics of neural networks.  This concerns especially invariant circles and saddle fixed points, which can appear to be stable in restrictions to some invariant subspaces of the 3D maps.  As an example, let us discuss the 1D maps shown in Fig.~\ref{fig15}. The map in Fig.~\ref{sfig25} depicts 
the transitioning behavior of the forward iterates of $\Delta \phi_{13}$ (modulo 1) towards a stable fixed point at $3/4$ on the $45^\circ$-line. First,  the iterates approach from above a phantom at $1/2$  on the $45^\circ$-line corresponding to the  saddle in the 3D phase space of the full map. Having lingered by, the phase point runs down to re-emerge at the top left corner and next at the right corner of the map. Finally, it  converges to the stable fixed point $ \Delta \phi_{13}^*=3/4$ from the right.  
Figure~\ref{sfig26} depicts the behavior of the same sequence in the $\Delta \phi_{14}$-projection that tends to the corresponding fixed point at $1/2$, shifted by a half period. In this projection, the coordinate of the saddle fixed point is $1$. 

\section{Discussion}

Many abnormal neurological phenomena are perturbations of normal mechanisms that govern behaviors such as movement. Repetitive motor behaviors are often hypothesized to correspond to rhythmogenesis in small networks of neurons that are able autonomously to generate or continue a variety of activity patterns without further external input. The detailed correspondence across these scales has not yet been made clear in any animal. But there is a growing consensus in the community of neurophysiologists and computational researchers that some basic structural and functional elements are likely shared by CPGs of both invertebrate and vertebrate animals. Before we can study the mechanisms of disorders at the level of individual neurons and CPG circuits in mammals, we therefore first seek to develop better tools and techniques in the context of much simpler animals. However, the ultimate aim of developing our tools and approach to understanding CPGs in lower animals is to make them applicable to studying the governing principles of neurological phenomena in higher animals, and so could potentially assist in treating neurological disorders associated with CPG arrhythmia. Our presentation here is intended as a tutorial guide that demonstrates the effectiveness of our analytical approach that connects exploratory mathematical models to experimental data in the context of known behavioral patterns.

In this pilot study, we focused on a biological CPG that has been linked to a specific swim motion in the sea slug, {\it Melibe}. This CPG robustly produces a phase-locked bursting pattern according to recent simultaneous voltage recordings from the four identified neurons. The goal of this modeling study was twofold: first, to identify specific components and their connections in this CPG; second, to get insight into how the connection `weights' in this network lead to robust production of the rhythmic patterns of self-sustained activity. Given the preliminary state of our knowledge about CPGs in the context of whole-animal behavior, we only aim to capture broad rhythmic properties of a small biophysical network without attempting to model each identified {\it Melibe} swim CPG neuron in precise detail. Thus, we avoid fitting microscopic details in our model such as a precise constitution of ionic currents, exact shapes and numbers of action potentials per bursts. Instead, we employ a generic Hodgkin-Huxley type model of endogenously bursting neurons that are qualitatively typical for invertebrates. This allows us to concentrate on explore a wide range of network configurations that might be responsible for the specific bursting patterns observed in the {\it Melibe} CPG.

We also intentionally construct and present our approach in a fashion that is analogous to the dynamic clamp technique used in neurophysiological experiments. Our technique involves the \emph{dynamic} removal, restoration and variance of (chemical) synaptic connections during simulation, which mimics the experimental techniques of drug-induced synaptic blockade, wash-out, etc. Restoring the chemical synapses during a simulation makes the CPG regain a bursting pattern with temporal characteristics such as phase lags, duty cycles, that depend on the connectivity strengths between the interneurons of the network.  

Due to the rhythmic nature of the bursting patterns, we employ Poincar\'e return maps defined on phases and phase lags between burst initiations in the interneurons. These maps allow us to study quantitative and qualitative properties of the stable rhythms and their corresponding attractor basins. We also exploit conventional knowledge about anti-phase bursting patterns to identify some basic requirements for plausible network configurations. For instance, reciprocal inhibition between a pair of neurons has long been known to produce anti-phase bursts in half center oscillator (HCO) configurations. We rely on a common, standing assumption in current neuroscience that the circuits of motor generation and control are modular in nature. Thus, the present theoretical challenge is how to understand the HCOs as building blocks that must be interconnected to produce single or multiple bursting patterns robustly, and what determines the stability and predominance of these rhythms. Our study is a step towards the dynamical foundation of this theory. We find that our model of the CPG reproduced, quite accurately, the available intracellular recordings from identified interneurons in the {\it Melibe} CPG. Furthermore, we find that the {\it Melibe} network can be interpreted to consist of two interconnected individual HCOs of two neurons each.

Furthermore, depending on strengths of unidirectional inhibition and excitation, we find that the individual HCOs may have different distributions of phase-locked states. This is a significant observation because, for example, inter-cellular recordings from the identified interneurons of the swim CPG of a similar sea slug, {\it Dendronotus}, indicate a  phase-locked state that is consistent with our model when it is configured using dominant, ipsilateral excitatory connections from one individual HCO to the other. In addition, as shown in the  ``Range of heterogeneity'' section,  the coupling strengths of the reciprocal inhibition within the HCOs have to be balanced in a certain ratio for the whole network to achieve the desired phase-locking.

We dissected the geometric organization of the simulation results into interactions between the building blocks of uncoupled and coupled HCOs. The relations between the phase lags helped us to link network architecture (configuration) with geometric organization of the solutions (model output). For instance, we show that each attractor of the network, whether it is a fixed point or an invariant circle, corresponds to either a phase-locked bursting pattern with distinct phase lags, or else to a bursting pattern with phase lags that vary periodically over the whole network period.  As it is unknown, {\it a priori}, whether the {\it Melibe} swim CPG is multifunctional for given set of parameters, one needs to sweep all possible initial phase distribution to reveal the existence of multiple attractors in the phase space of the corresponding return map.  Through the use of decoupled HCOs  we were able to explain significant details of the 3D phase portraits such as convergence rates and the occurrence of designated convergence routes to attractors of the phase-lag map.

To study the known robustness of our network, we use an ensemble of computational tools that allow for the reduction of observable voltage dynamics to low-dimensional return maps for phase lags between burst initiations in the interneurons. In particular, we reduce the bursting dynamics of the 4-cell network, represented in full by a 12-dimensional system of ODEs, to a 3-dimensional return map for the phase lags between either the endogenously bursting interneurons or between bursting HCOs. We use a ``top-down'' approach in which we systematically examine the properties of the phase-lag maps that we abstracted from the 12D system rather than exploring the global dynamics of that full system directly. With this approach, we identified that both contralateral feed-forward inhibition and ipsilateral backward excitation are needed for the network to stabilize the bursting pattern against small perturbations. A certain balance of the synaptic strengths is also required to maintain the phase-locked state within a reasonable range in the parameter space of the CPG network. 

Another strong working assumption in this study is that a CPG is composed of (nearly) identical elements --- interneurons or bursting HCOs --- which are interconnected through chemical synapses and gap junctions of equal conductivity. Due to alterations in the reciprocal wiring, such a homogeneous CPG can be adapted to become dedicated to a single rhythm or multifunctional. We might presume that, through iterative processes of learning and evolution, a real CPG might develop a heterogeneous structure as specific connections become stronger or weaker, so that it can become better adapted to performing specific functions in specific animals. Certainly, we are all aware of examples where, through learning and exercise, mammalian motor systems become ``multifunctional'' and are able to quickly transition between several dynamic functions on demand: for instance, the diverse swimming styles that have been cultivated by humans, including the in-phase breaststroke and butterfly, and the anti-phase crawl and backstroke. For now, we can only hypothesize that there is a multifunctional, and presumably heterogeneous, CPG network underlying these specific swimming rhythms that determines the phase relationships between rhythmic muscle control signals.

In general, our insights allow us to predict both quantitative and qualitative transformations of the observed patterns whenever the network configurations are altered. The nature of these transformations provides a set of novel hypotheses for biophysical mechanisms about the control and modulation of rhythmic activity. A powerful aspect to our analytical technique is that it does not require knowledge of the equations that model the system. Thus, we believe that have further developed a universal approach to studying both detailed and phenomenological models of bursting networks that is also applicable to a variety of rhythmic biological phenomena beyond motor control. 

Even the real {\it Melibe} swim CPG is, of course, much more complex than our specific model that is based on the existing, preliminary experimental data. There is great room for improvement in the model by incorporating other biological features into it. There are many open questions that could be addressed by more detailed modeling, such as whether the individual models are natural bursters with distinct duty cycles, or whether they spike tonically and can only become network bursters episodically when under the influence of external drive from other pre-synaptic interneurons in the CPG. This is a challenging question, both phenomenologically and computationally. In future work based upon the framework of this study, we plan to address such questions with more realistic models of inhibitory-excitatory CPG configurations, including ones comprised of three and more HCOs.
 
\section{Acknowledgments} This work is funded by NSF grant DMS-1009591, RFFI Grant No. 08-01-00083, MESRF project 14.740.11.0919, as well as GSU Brain and Behaviors Initiative. Research of D. Allen and J. Youker was supported by NSF REU. We would like to thank A. Kelley, J. Schwabedal, R. Clewley, A. Sakurai and P. Katz for helpful discussions and comments. We are also grateful to J. Wojcik and R. Clewley for guidance  on the PyDSTool package \cite{PyDSTool} used for  network simulations.

\section{Appendix} \label{appendix}

\subsection{Leech heart interneuron model}
This networked model is based on the dynamics of the fast sodium current, $I_{\rm Na}$, the slow potassium current, $I_{\rm K2}$, and an ohmic leak current, $I_{\rm L}$
is given by \cite{SGB08,WCS11}:
\begin{equation}
\begin{array}{l}
C\, \displaystyle  \frac{dV}{dt}  =  -I_{\mathrm{Na}}(V)-I_{\mathrm{K2}}(V)-I_{\mathrm{L}}(V)-I_{\mathrm{app}} - I_{\rm syn} , \\
I_{\mathrm{Na}}={\bar g}_{\mathrm{Na}}\,n^3\,h\,(V-E_{\mathrm{Na}}),
\quad  n=n^\infty(V), \\
I_{\mathrm{K2}}={\bar g}_{\mathrm{K2}}\,m^2(V-E_{\mathrm{K}}), \quad I_{L}=\bar g_{\mathrm{L}}\,(V-E_{\mathrm{L}}),    \\
\tau_{\mathrm{Na}}\, \displaystyle   \frac{dh}{dt} = h^\infty(V)-h,
 \quad
\tau_{\mathrm{K2}} \displaystyle   \frac{dm}{dt} = m^\infty(V)-m.
\end{array}
\label{leech1}
\end{equation}
Here,  $V$ is the membrane potential, $n$ and $h$ are the gating variables for sodium channels, which activate (instantaneously) and inactivate, respectively, as the membrane potential depolarizes; $m$ is the gating variable for the potassium channel that activates slowly as the membrane potential hyperpolarizes. An applied current, $I_{\rm app}=0$, through the paper unless indicated otherwise. The time constants for the gating variables, maximum conductances and reversal potentials for all the channels and leak current, and the membrane capacitance are set as follows:
\begin{equation}
\begin{array}{lcr}
\tau_{\rm Na}=0.0405~{\rm sec},        &	{\bar g}_{\rm Na}=200~{\rm nS},  	&	E_{\rm Na}=0.045~{\rm V}, \nonumber \\
\tau_{\rm{K2}}=0.25~{\rm sec},         &	{\bar g}_{\rm K2}=30~{\rm nS},          & 	E_{\rm K}= -0.070~{\rm V},  \nonumber \\
C=0.5~{\rm nF},				    &		{\bar g}_{\rm L}=8~{\rm nS},  		&	E_{\mathrm{L}}=-0.046~{\rm V}. \nonumber
\end{array}
\label{leech2}
\end{equation}
The steady state values of the gating variables are given by the following Boltzmann functions:
 \begin{equation}
\begin{array}{rcl}
 n^\infty(V)&=&[1+\exp(-150(V+0.0305))]^{-1},  \nonumber \\
 h^\infty(V)&=&[1+\exp(500(V+0.0333))]^{-1},  \nonumber \\
\quad m^\infty(V)&=&[1+\exp{(-83(V+0.018+\mathrm{V^{shift}_{K2}}))}]^{-1}, \nonumber
\end{array}
\label{leech3}
\end{equation}
with $\mathrm{V^{shift}_{K2}}=-0.02181$V; this  bifurcation parameter controls the number of spikes per burst. The currents through fast, non-delayed, chemical synapses are modeled using the fast threshold modulation paradigm as follows \cite{FTM}:
 \begin{equation}
 \begin{array}{rcl}
I_{\rm syn} &=& g_{\rm syn} (E_{\rm syn} -V_{\mathrm{post}}) \Gamma (V_{\mathrm{pre}}-\Theta_{\rm syn}),\\
\Gamma(V_{\rm pre}-\Theta_{\rm syn}) &=& 1/[1+{\rm exp}\{-1000(V_{\mathrm{pre}}-\Theta_{\rm syn})\}];
\end{array}
\end{equation}
here $V_{\mathrm{post}}$  and $V_{\mathrm{pre}}$ are the voltages of the post- and the pre-synaptic interneurons; the synaptic
threshold $\Theta_{\rm syn}=-0.03$V is chosen so that every  spike within a burst in the pre-synaptic neuron crosses $\Theta_{\rm syn}$. This implies that the synaptic current, $I_{\rm syn}$, is initiated as soon as $V_{\mathrm{pre}}$ {exceeds the
synaptic threshold. The type, inhibitory or excitatory, of the FTM synapse is determined by the level of the reversal potential, $E_{\rm syn}$,
in the post-synaptic neuron.
In the inhibitory case, it is set as $E_{\rm syn}=-0.0625$V so that $V_{\rm post}(t)> E_{\rm syn}$. In the excitatory case {the level of $E_{\rm syn}$ is raised to zero to guarantee that $\langle V_{\rm post}(t) \rangle$ remains,  below the reversal potential
on average, over the period of the bursting interneuron. We point out specifically that our previous studies of network dynamics revealed that  alternative models of fast chemical synapses, using the  alpha function and  detailed dynamical representation, have no contrast effect on  interactions between the coupled interneurons \cite{pre2012}.

\subsection{Phase-lags and HCO periods}

The delays between burst initiations in the reference  interneuron~1 and other three are given by $\tau_{1j}$, and the corresponding times are given by $t_{j}$, and recurrent times (network period) by $ \tilde T_{j}$ ($T_{j}$). The superscripts stand for bursting cycle numbers. Then,  the phase-lags are defined as the following:
\begin{equation}
\Delta \phi^{(n)}_{1j} = \frac{t^{(n)}_{j}-t^{(n)}_{1}}{t^{(n)}_{1}-t^{(n-1)}_{1}}= \frac{\tau^{(n)}_{1j}}{ \tilde T^{(n)}_{1}}, \quad j=2,3,4.
\label{dph}
\end{equation}
After the phase-locked state is achieved the difference between subsequent phase-lags does not change: 
\begin{equation}
\Delta \phi^{(n+1)}_{1j}-\Delta \phi^{(n)}_{1j} = 0, \quad \mbox{and} \quad \frac{\tau^{(n+1)}_{1j}}{\tilde T^{(n+1)}_{1}}-\frac{\tau^{(n)}_{1j}}{\tilde T^{(n)}_{1}} = 0.
\label{ddph}
\end{equation}
The equality, $T^{(n+1)}_{1} = T^{(n)}_{1}$,  means that HCO1 maintains a constant period $T_1$. Then the above conditions can be reduced to those on the following ratios of the periods between both HCOs:
\begin{equation}
\begin{array}{rcl}
\Delta \phi^{(n+1)}_{1j}-\Delta \phi^{(n)}_{1j} &=& \displaystyle \frac{\tau^{(n+1)}_{1j}}{T_{1}}-\frac{\tau^{(n)}_{1j}}{T_{1}}=\frac{ \left((t^{(n+1)}_{j}-t^{(n+1)}_{1}) - (t^{(n)}_{j}-t^{(n)}_{1}) \right)} {T_1}\\~\\
\displaystyle
 &=& \displaystyle  \frac{ \left ((t^{(n+1)}_{j}-t^{(n)}_{j})-(t^{(n+1)}_{1}-t^{(n)}_{1} )\right )} {T_1}
 =\frac{T^{(n+1)}_j-T_1}{T_1} = \frac{T^{(n+1)}_{j}}{T_{1}}-1.\\
\end{array}\label{deldelphi}
\end{equation}
Whenever the phase-locked state is achieved, HCO2 has the period  of HCO1.  On the other hand, if $T^{(n+1)}_{1} \neq T^{(n)}_{1}$, then the delays $\tau^{(n)}_{1j}$ and $\tau^{(n+1)}_{1j}$ will change proportionally too: 
\begin{equation}
\Delta \phi^{(n+1)}_{1j}-\Delta \phi^{(n)}_{1j} = \frac{\tau^{(n+1)}_{1j}}{T^{(n+1)}_{1}} - \frac{\tau^{(n)}_{1j}}{T^{(n)}_{1}}= 0, \quad \mbox{or} \quad 
\frac{T^{(n)}_1}{T^{(n+1)}_1} = \frac{\tau^{(n)}_{1j}}{\tau^{(n+1)}_{1j}}.
\label{deldelphi}
\end{equation}

\subsection{Numerical methods} All numerical simulations and phase analysis were performed using the PyDSTool dynamical systems software environment \cite{PyDSTool}. 
Each sequence of phase lags $\{ \Delta \phi_{1j}^{(n)} \}$ plotted in Fig.~1 begins from an initial lag $(\Delta \phi_{13}^{(0)})$, which is the difference in phases
measured relative to the recurrence time of cell 1 every time its voltage increases to a threshold $\Theta_{\rm th} = -40$~mV. $\Theta_{\rm th}$ marks the beginning of the spiking segment of a burst.
As that recurrence time is unknown a priori due to interactions of the cells, we estimate it,  up to first order, as a fraction of the period $T_\mathrm{synch}$ of the synchronous bursting orbit (or that in the individual models) by selecting guess values $(\Delta \phi_{1j}^{\star})$. The synchronous solution corresponds to $\Delta \phi_{1j} = 0$.
By identifying  $t=0$ at the moment when $V_1=\Theta_{\rm th}$ with $\phi=0$, we can parameterize this solution by time ($0\leq t < T_\mathrm{synch}$) or by the phase lag $(0 \leq \Delta \phi < 1)$. For weak coupling and small lags, the recurrence time is close to $T_\mathrm{synch}$, and $(\Delta \phi_{1j}^{\star}) \approx (\Delta \phi_{1j}^{(0)})$. We use the following algorithm to distribute the true initial lags uniformly on a $40 \times 40$ square grid covering the the unit cube (torus), which is the phase space of the phase-lag network.

We initialize the state of cell 1 at $t=0$ from the point $(V^0, n^0, h^0)$ of the synchronous solution, and  next
%when $V_1 = \Theta_{\rm th}$.  
create the distibution of the initial phase-lagged states in the simulation by suppressing the other cells for durations $t=\Delta \phi_{1j}^{0} T_\mathrm{synch}$, respectively. On release, the cells are initialized with the same state $(V^0, n^0, h^0)$. We begin recording the sequence of phase lags between the cells 2--4 and the reference cell 1 on the second cycle after coupling has adjusted the network period away from $T_\mathrm{synch}$. In the case of stronger coupling, where the gap between  $T_\mathrm{synch}$ and the first recurrence time for cell 1 widens, we retroactively adjust initial phases using a  ``shooting'' algorithm to make the initial phase lags sufficiently close to uniformly distributed positions on the square grid.

\end{document}